\documentclass[runningheads]{llncs}
\usepackage[T1]{fontenc}
\usepackage{lmodern}
\usepackage{microtype}
\usepackage{hyperref}
\usepackage{url}
\usepackage{booktabs}
\usepackage{cite}
\usepackage{lineno}
\usepackage{caption}

\usepackage[normalem]{ulem}

\usepackage{amsmath}
\usepackage{amssymb}
\usepackage{algorithm}
\usepackage{algorithmicx}
\usepackage{algpseudocode}
\usepackage{float}
\usepackage{listings}
\usepackage{graphicx}
\usepackage{tikz}
\usepackage{makecell}
\usepackage{multirow}
\usepackage{tcolorbox}
\usepackage{enumitem}
\usepackage{wrapfig}

\usepackage{color}

\tcbuselibrary{listings,skins,breakable}
\lstdefinelanguage{tlaplus}{
  keywords={EXTENDS, THEOREM, ASSUME, PROVE, NEW, OBVIOUS, BY, MODULE},
  keywordstyle=\color{darkblue}\bfseries,
  ndkeywords={Int, TRUE, FALSE},
  ndkeywordstyle=\color{purple}\bfseries,
  identifierstyle=\color{black},
  sensitive=true,
  comment=[l]{//},
  commentstyle=\color{green!60!black}\itshape,
  stringstyle=\color{red},
  morestring=[b]',
  morestring=[b]"
}

\lstdefinestyle{tlaplusfancy}{
  language=tlaplus,
  basicstyle=\ttfamily\scriptsize,
  numberstyle=\tiny\color{gray},
  numbers=left,
  numbersep=5pt,
  backgroundcolor=\color{gray!5},
  showspaces=false,
  showstringspaces=false,
  showtabs=false,
  frame=single,
  frameround=tttt,
  framesep=1pt,
  rulecolor=\color{black!30},
  captionpos=b,
  breaklines=true,
  breakatwhitespace=true,
  title=\lstname,
  belowskip=0pt,
  escapeinside={<@}{@>},
  deletestring=[b]',
  morecomment=[l]{\\*},
  commentstyle=\color{black!50}\itshape,
  morekeywords={VARIABLE, VARIABLES, LOCAL, INSTANCE, CONSTANT, CONSTANTS, RECURSIVE,
                UNION, SUBSET, DOMAIN, ENABLED, UNCHANGED, EXCEPT, LET, IN, CASE,
                OTHER, IF, THEN, ELSE, SF, WF, DEF},
  literate={\\A}{{$\forall$}}1 
           {\\E}{{$\exists$}}1
           {=>}{{$\Rightarrow$}}2 
           {\\in}{{$\in$}}1
           {/\\}{{$\land$}}1
           {\\/}{{$\lor$}}1
           {\\cap}{{$\cap$}}1
           {\\cup}{{$\cup$}}1
           {\\intersect}{{$\cap$}}1
           {\\union}{{$\cup$}}1
           {\\subset}{{$\subset$}}1
           {\\supset}{{$\supset$}}1
           {\\emptyset}{{$\emptyset$}}1
           {/=}{{$\neq$}}1
           {\\neq}{{$\neq$}}1
           {\\neg}{{$\neg$}}1
           {\\land}{{$\land$}}1
           {\\lor}{{$\lor$}}1
           {~}{{$\sim$}}1
           {\\equiv}{{$\equiv$}}1
           {\\cdot}{{$\cdot$}}1
           {\\o}{{$\circ$}}1
           {\\circ}{{$\circ$}}1
           {\\Box}{{$\Box$}}1
           {\\Diamond}{{$\Diamond$}}1
           {[]}{{{$\Box$}}}1
           {<>}{{{$\Diamond$}}}1
           {~}{{$\neg$}}1
           {\#}{{$\neq$}}2
           {>=}{{$\geq$}}2
           {<=}{{$\leq$}}2
}

\definecolor{systemBg}{RGB}{240, 242, 245}
\definecolor{systemBorder}{RGB}{200, 210, 220}
\definecolor{userBg}{RGB}{235, 245, 250}
\definecolor{userBorder}{RGB}{180, 210, 230}
\definecolor{promptComment}{RGB}{100, 100, 100}
\definecolor{promptKeyword}{RGB}{0, 100, 150}
\definecolor{promptString}{RGB}{160, 70, 160}

\lstdefinestyle{systemprompt}{
    backgroundcolor=\color{systemBg},
    frame=single,
    framesep=8pt,
    framexleftmargin=12pt,
    framexrightmargin=6pt,
    framextopmargin=6pt,
    framexbottommargin=6pt,
    rulecolor=\color{systemBorder},
    basicstyle=\small\ttfamily,
    commentstyle=\color{promptComment}\itshape,
    keywordstyle=\color{promptKeyword}\bfseries,
    stringstyle=\color{promptString},
    breaklines=true,
    breakatwhitespace=false,
    captionpos=b,
    keepspaces=true,
    showspaces=false,
    showstringspaces=false,
    showtabs=false,
    tabsize=2,
    title={\textbf{System Prompt}},
    morekeywords={INSTRUCTION, CONTEXT, EXAMPLES, TASK, NOTE},
}

\lstdefinestyle{userprompt}{
    backgroundcolor=\color{userBg},
    frame=single,
    framesep=8pt,
    framexleftmargin=12pt,
    framexrightmargin=6pt,
    framextopmargin=6pt,
    framexbottommargin=6pt,
    rulecolor=\color{userBorder},
    basicstyle=\small\ttfamily,
    commentstyle=\color{promptComment}\itshape,
    keywordstyle=\color{promptKeyword}\bfseries,
    stringstyle=\color{promptString},
    breaklines=true,
    breakatwhitespace=false,
    captionpos=b,
    keepspaces=true,
    showspaces=false,
    showstringspaces=false,
    showtabs=false,
    tabsize=2,
    title={\textbf{User Prompt}},
    morekeywords={QUERY, INPUT, QUESTION, PROBLEM},
}

\usetikzlibrary{arrows.meta,positioning,shapes,arrows,shadows,fit,calc,backgrounds}

\tikzset{
    goal/.style={
        draw,
        rounded corners=5pt,
        fill=black!5,
        text width=3.5cm,
        align=center,
        minimum height=1cm,
        font=\small\sffamily,
        drop shadow={shadow xshift=0.3mm, shadow yshift=-0.3mm}
    },
    lemma/.style={
        draw,
        rounded corners=5pt,
        minimum width=2.5cm,
        minimum height=0.7cm,
        text width=2.4cm,
        align=center,
        font=\small\sffamily,
        drop shadow={shadow xshift=0.2mm, shadow yshift=-0.2mm}
    },
    obvious/.style={
        draw,
        rounded corners=3pt,
        minimum width=1.5cm,
        minimum height=0.5cm,
        align=center,
        font=\scriptsize\sffamily,
        drop shadow={shadow xshift=0.1mm, shadow yshift=-0.1mm}
    },
    factorform/.style={lemma, fill=blue!10, text=black},
    solvefactors/.style={lemma, fill=orange!15, text=black},
    connection/.style={
        thick,
        -latex,
        shorten >=1pt,
        shorten <=1pt
    }
}

\definecolor{darkblue}{rgb}{0, 0, 0.5}
\hypersetup{colorlinks=true, citecolor=darkblue, linkcolor=darkblue, urlcolor=darkblue}

\newcommand{\extendedVersion}{}

\newcommand{\terminalProof}{\mathit{autoProof}}

\newcommand{\wordproofs}{\mathit{proofs}}
\newcommand{\wordproof}{\mathit{proof}}
\newcommand{\wordsubClaims}{\mathit{subClaims}}

\newcommand{\wordclaim}{\mathit{claim}}
\newcommand{\wordcontext}{\mathit{context}}
\newcommand{\wordgoal}{\mathit{goal}}

\newcommand{\worddecompositionValid}{\mathit{decompositionValid}}

\newcommand{\tlaplus}{\texttt{TLA}^+}
\newcommand{\TLAPM}{\texttt{TLAPS} }

\title{Towards Language Model Guided $\text{TLA}^{+}$ Proof Automation}

\date{}

\begin{document}

    \author{Yuhao Zhou \and Stavros Tripakis}
    \authorrunning{Yuhao Zhou and Stavros Tripakis}
    \institute{Northeastern University, {Boston}, {MA}}

    \maketitle            

    \begin{abstract}
    Formal theorem proving with $\texttt{TLA}^{+}$ provides rigorous guarantees for system specifications, but constructing proofs requires substantial expertise and effort. While large language models have shown promise in automating proofs for tactic-based theorem provers like Lean, applying these approaches directly to $\texttt{TLA}^{+}$ faces significant challenges due to the hierarchical proof structure of the $\texttt{TLA}^{+}$ proof system. We present a prompt-based approach that leverages LLMs to guide hierarchical decomposition of complex proof obligations into simpler sub-claims, while relying on symbolic provers for verification. Our key insight is to constrain LLMs to generate normalized claim decompositions rather than complete proofs, significantly reducing syntax errors. 
    We also introduce a benchmark suite of 119 theorems adapted from (1) established mathematical collections and (2) inductive proofs of distributed protocols. Our approach consistently outperforms baseline methods across the benchmark suite.
    \keywords{Formal Verification \and Theorem Proving \and Large Language Models \and $\text{TLA}^{+}$}
    \end{abstract}

    \section{Introduction}
    \label{sec:introduction}
    Formal verification plays a crucial role in ensuring the correctness of critical systems, particularly distributed systems where subtle errors can have severe consequences. As systems grow in complexity and become more interconnected, the need for rigorous verification methods becomes increasingly vital. The $\tlaplus$ specification language~\cite{lamport2002tlaplus} has emerged as a powerful framework for modeling and verifying such systems, with significant adoption in companies like Amazon, Intel, and Microsoft~\cite{newcombe2014amazon_why_tla,beers2008intel_use_tla,hackett2023azure_tla,NewcombeAmazon2015}. 
Despite its effectiveness, constructing formal proofs in $\tlaplus$ remains time-consuming and requires substantial expertise, creating a bottleneck in the verification process~\cite{schultzOPODIS2021,SchultzDardikTripakisCPP2022}.

Proof automation is fundamentally challenging: the underlying problem of proving theorems in expressive logics is undecidable~\cite{church1936unsolvable,turing1936computable} and state-of-the-art provers still require substantial human guidance for complex proofs~\cite{Rocq,moura2021lean4}. 
Therefore, any progress in techniques that assist or automate proof construction represents a significant opportunity to lower the barrier to formal verification, by making it more practical and scalable.

Recent advances in Large Language Models (LLMs) have shown promise in automating formal theorem proving tasks, particularly in {\em tactic-based} theorem provers like Lean~\cite{yang2024leandojo} and Rocq (previously known as Coq)~\cite{thakur2023corpaInContextTheoremProving}.
These approaches leverage LLMs' capabilities to generate sequences of proof tactics that incrementally transform proof states toward the goal.
However, $\tlaplus$ employs a fundamentally different, {\em hierarchical} proof structure. Unlike Lean and Rocq, which sequentially transform proof states by tactics, $\tlaplus$ proofs are organized as trees of claims and sub-claims. For example, while a tactic-based proof consists of a sequence of transformations (e.g., `expand definition, apply distributive property, simplify'), a $\tlaplus$ proof introduces intermediate claims that collectively establish the goal. This distinction is illustrated by the proofs of theorem \texttt{T1} in Figure \ref{fig:prover_diff_tlaps_example} ($\tlaplus$) and Figure \ref{fig:prover_diff_lean4_example} (Lean).

\begin{figure}[t]
    \centering
    \begin{minipage}[t]{0.46\textwidth}
        \centering
        \begin{minipage}[t]{\textwidth}
            \centering
            \setlength{\abovecaptionskip}{0pt}
            \begin{lstlisting}[frame=single,style=tlaplusfancy]
-------- MODULE sums_even -------
EXTENDS Naturals, TLAPS

Even(x) == x % 2 = 0

THEOREM L0 == ASSUME NEW x \in Nat PROVE Even(x + x) = Even(x * 2) 
OBVIOUS

THEOREM L1 == ASSUME NEW x \in Nat PROVE Even(x * 2) = ((x * 2) % 2 = 0)
BY DEF Even

THEOREM T1 == ASSUME NEW x \in Nat PROVE Even(x + x)
BY L0, L1 DEF Even
=======================\end{lstlisting}
            \caption{Theorem \texttt{T1} proven in $\tlaplus$.} 
            \label{fig:prover_diff_tlaps_example}
        \end{minipage}
        \vfill
    \end{minipage}
    \hfill
    \begin{minipage}[t]{0.46\textwidth}
        \centering
        \setlength{\abovecaptionskip}{0pt}
        \begin{lstlisting}[frame=single,style=tlaplusfancy]
import Mathlib.Tactic.Ring   

def even (x : Nat) : Prop := x % 2 = 0

theorem T1 : \A x : Nat, even (x+x) := by
    intro x
    ring_nf
    dsimp [even]
    simp\end{lstlisting}
        \caption{The same theorem \texttt{T1} of Figure~\ref{fig:prover_diff_tlaps_example}, now proven in Lean. The Lean proof consists of a sequence of {\em tactics} (lines 6-9) that transform the proof state to solve the goal.}
        \label{fig:prover_diff_lean4_example}
    \end{minipage}
\end{figure}

Additionally, while systems like Lean and Rocq have extensive libraries of formalized proofs that can serve as training data and benchmarks for machine learning approaches, $\tlaplus$ lacks comparable datasets, creating a significant challenge for developing and evaluating learning-based proof automation.

In this paper, we present a language-model based approach to automating $\tlaplus$ proof generation. Our method, called
{\em Language Model Guided Proof Automation} (LMGPA), accommodates the hierarchical structure of $\tlaplus$ proofs through a recursive decomposition strategy. This approach guides LLMs to recursively break down complex claims into simpler sub-claims that can be independently verified, mirroring the natural structure of $\tlaplus$ proofs. Our system verifies each decomposition step, providing feedback to the LLM when necessary, and recursively applies the same process to each sub-claim until all claims can be verified by backend provers.

    \section{Preliminaries and Problem Statement}
    \label{sec:prelim_problem_statement}
    {\bf $\tlaplus$ and $\tlaplus$ Proof System}\quad
$\tlaplus$ is a formal specification language~\cite{lamport2002tlaplus}
designed for specifying and verifying properties of complex systems and algorithms, particularly distributed systems and concurrent algorithms. 
It has been adopted in both academia and industry, with companies such as Amazon, Microsoft, and Intel successfully applying it to verify critical systems and protocols~\cite{newcombe2014amazon_why_tla, hackett2023azure_tla,konnov2022specification_in_tla, beers2008intel_use_tla}.
$\tlaplus$ is supported by the $\tlaplus$ Foundation -- see \url{https://foundation.tlapl.us/}.

As a language grounded in mathematical logic, $\tlaplus$ enables not only precise specification but also rigorous verification through model checking~\cite{yu1999TLC} and theorem proving. 
While model checking is an essential formal verification method~\cite{ClarkeGrumbergPeledBook,Baier2008PMC,handbookMC}, in the industry it is typically used for finding error traces quickly, and for verifying correctness of finite-state systems or bounded instances of infinite-state systems. In this paper, we focus on formal theorem proving, which allows to prove correctness of unbounded/infinite systems. 
Theorem proving for $\tlaplus$ is implemented in the $\tlaplus$ 
Proof System ($\TLAPM$)~\cite{chaudhuri2010tlapm}, which serves as a bridge between human-written specifications and automated verification tools.
$\TLAPM$ translates $\tlaplus$ specifications and proofs into forms supported by backend provers like Z3~\cite{de2008z3}, Zenon~\cite{bonichon2007zenon}, and Isabelle~\cite{paulson1994isabelle,isabelle-hol-Nipkow-Paulson-Wenzel}. 

The proving approach in $\tlaplus$ represents a distinct paradigm when compared to other prominent formal theorem provers, particularly in how proofs are structured and developed.
In what follows, we discuss the most important differences.

{\bf \TLAPM vs Tactic-based Interactive Theorem Provers}\quad
In the landscape of formal theorem proving, many popular \textit{Interactive Theorem Provers} (ITPs) such as Lean~\cite{moura2021lean4}, Rocq~\cite{Rocq}, and Isabelle/HOL~\cite{isabelle-hol-Nipkow-Paulson-Wenzel} support a tactic-based approach to proof construction.\footnote{
Tactical proofs are not the only style these provers support. For example, 
Lean also supports calculational proofs.
Isabelle also supports hierarchically structured proofs similar to TLAPS; in fact, Isabelle is one of the backend provers used by $\TLAPM$.}
In the tactical style of proving, 
machine-checkable formal proofs are expressed as sequences of \textit{tactics}---commands that systematically transform the proof state. Users guide the proof development by iteratively applying these tactics, effectively directing the prover through the proving process.
The Lean proof in Figure \ref{fig:prover_diff_lean4_example} illustrates this: the proof of \texttt{T1} is a sequence of tactics (lines 6-9) like \texttt{intro}, \texttt{ring\_nf}, and \texttt{simp} that manipulate and solve the proof goal.

The proof methodology in $\tlaplus$, however, follows a fundamentally different structure. Rather than tactical transformations, $\tlaplus$ proofs are organized hierarchically---users establish complex claims by identifying and introducing intermediate sub-claims. 
For instance, the $\tlaplus$ proof in Figure \ref{fig:prover_diff_tlaps_example} demonstrates this structure, the explicit intermediate sub-claims \texttt{L0} and \texttt{L1} collectively establish the goal \texttt{T1}.
This hierarchical proof continues growing until the entire proof is directly machine-checkable by backend provers. This methodological distinction has significant implications for how proofs are developed, understood, and potentially automated within the $\TLAPM$.

It is important to note that the proofs in Figures \ref{fig:prover_diff_tlaps_example} and \ref{fig:prover_diff_lean4_example} are presented purely for illustrative purposes to highlight this methodological difference. More direct or idiomatic proofs of \texttt{T1} exist in both systems. While the theorem \texttt{T1} is adapted from the $\tlaplus$ example repository~\cite{tlaplus_examples}, the $\tlaplus$ proof structure was intentionally modified to explicitly demonstrate the differences between hierarchical and tactic-based proving approaches.

\begin{figure}[th]
    \centering
    \begin{minipage}[t]{0.46\textwidth}
        \centering
        \begin{minipage}[t]{\textwidth}
            \centering
            \setlength{\abovecaptionskip}{0pt}
            \begin{lstlisting}[frame=single,style=tlaplusfancy]
-------- MODULE amc12a_2015_p10 -------
EXTENDS Integers, TLAPS
            
THEOREM Main == 
    \A x, y \in Int: (0 < y) /\ (y < x) /\ (x + y + (x * y) = 80) => (x = 26)
=======================\end{lstlisting}
            \caption{An example theorem represented in $\tlaplus$ as input to our proof generation system.}
            \label{fig:example_input}
        \end{minipage}
        \vfill
        \vspace{4mm}
        \begin{minipage}[t]{\textwidth}
            \centering
            \setlength{\abovecaptionskip}{5pt}
            \resizebox{\textwidth}{!}{\begin{tikzpicture}[node distance=1.0cm, transform shape, scale=1.2]
    \node[goal] (goal) {Main\\
        \scriptsize{$\forall x, y \in \mathbb{Z}: (0 < y) \wedge$\\$(y < x) \wedge (x + y + xy = 80)$\\$\Rightarrow (x = 26)$}};
    
    \node[solvefactors, below left=0.8cm and 0.0cm of goal] (solvefactors) {SolveFactors};
    \node[factorform, below right=0.8cm and 0.0cm of goal] (factorform) {FactorForm};
    
    \node[obvious, below=0.8cm of solvefactors] (obvious1) {OBVIOUS};
    \node[obvious, below=0.8cm of factorform] (obvious2) {OBVIOUS};
    
    \draw[connection, orange!70!black] (goal.south) -- ++(0,-0.3) -| (solvefactors.north);
    \draw[connection, blue!70!black] (goal.south) -- ++(0,-0.3) -| (factorform.north);
    
    \draw[connection, orange!70!black] (solvefactors) -- (obvious1);
    \draw[connection, blue!70!black] (factorform) -- (obvious2);
\end{tikzpicture}}
            \caption{Visualization of the proof tree for the proof in Figure~\ref{fig:expected_example_output}.}
            \label{fig:proof_tree_example}
        \end{minipage}
    \end{minipage}
    \hfill
    \begin{minipage}[t]{0.46\textwidth}
        \centering
        \setlength{\abovecaptionskip}{0pt}
        \begin{lstlisting}[frame=single,style=tlaplusfancy]
-------- MODULE amc12a_2015_p10 -------
EXTENDS Integers, TLAPS

THEOREM FactorForm == 
    ASSUME  NEW x \in Int, NEW y \in Int,
            0 < y, y < x, 
            x + y + (x * y) = 80 
    PROVE   (x + 1) * (y + 1) = 81
OBVIOUS

THEOREM SolveFactors == 
    ASSUME  NEW x \in Int, NEW y \in Int,
            0 < y, y < x, 
            (x + 1) * (y + 1) = 81 
    PROVE   x = 26
OBVIOUS

THEOREM Main == 
    \A x, y \in Int: (0 < y) /\ (y < x) /\ (x + y + (x * y) = 80) => (x = 26)
BY FactorForm, SolveFactors
=======================\end{lstlisting}
        \caption{Complete $\tlaplus$ proof of the theorem in Figure~\ref{fig:example_input} (the entire proof was generated fully automatically by our system).}
        \label{fig:expected_example_output}
    \end{minipage}
\end{figure}

{\bf $\tlaplus$ Proof Structure}\quad
In $\tlaplus$, a proof is a hierarchical arrangement of claims, where each claim represents a theorem to prove. To illustrate this structure, we refer to the examples in Figures~\ref{fig:example_input} and~\ref{fig:expected_example_output}, which demonstrate a theorem and its corresponding proof in $\tlaplus$. Using these examples as reference points, we now define the key terminology used throughout this paper:
\begin{itemize}[leftmargin=*]
    \item A {\it claim} is a boolean-valued expression written in $\tlaplus$. For instance, line 5 in Figure~\ref{fig:example_input} (which is identical to line 19 in Figure~\ref{fig:expected_example_output}) is a claim. Lines 5-8 of Figure~\ref{fig:expected_example_output} collectively form another claim.
    
    \item A {\it goal} is a specific claim that requires proof, representing the theorem or lemma of interest. In our example, the \texttt{Main} claim serves as the goal.
    
    \item {\it Context} is the collection of definitions and assumptions that provide the logical foundation for the goal. This includes imported modules such as the \texttt{Integers} module in Figure~\ref{fig:example_input}, which provides the definition of \texttt{Int}.
    
    \item A {\it proof obligation} is a tuple of a context and a goal.

    \item A core construct in $\tlaplus$ proofs is the \texttt{ASSUME-PROVE} structure, as seen in \texttt{FactorForm} and \texttt{SolveFactors} in Figure~\ref{fig:expected_example_output}. These are interpreted as logical implications where \texttt{ASSUME} $F$ \texttt{PROVE} $G$ means $\vdash F \Rightarrow G$, i.e., prove that $F$ implies $G$.
\end{itemize}
While $\tlaplus$ offers a rich and expressive proof language with multiple approaches to establishing claims, this paper focuses on a specific subset of proof structures for clarity and tractability. Specifically, we consider proofs that follow the pattern demonstrated in Figure~\ref{fig:expected_example_output}, where a claim may be associated with one of the following:
\begin{itemize}[leftmargin=*]
    \item An \textit{auto proof}, where the claim can be directly verified by backend provers. In $\TLAPM$, auto proofs use either the keyword \texttt{OBVIOUS} or the form \texttt{BY DEF} with a list of definitions to unfold. For example, the proofs of theorems \texttt{FactorForm} and \texttt{SolveFactors} in Figure~\ref{fig:expected_example_output} use \texttt{OBVIOUS}, meaning they can be verified directly without unfolding any definitions. On the other hand, the proof of theorem \texttt{Main} is not an auto proof because it references other theorems.

    \item 
    A proof by sub-claims, where the parent claim is established by a set of sub-claims. In $\TLAPM$, this is expressed using the \texttt{BY} keyword followed by a list of the sub-claims.
	In Figure~\ref{fig:expected_example_output}, the parent claim \texttt{Main} is supported by two sub-claims: \texttt{FactorForm} and \texttt{SolveFactors}, which together provide a justification. Formally, if the parent claim asserts $F \Rightarrow G$, and it is supported by sub-claims $A$ and $B$, then we are submitting to the solver the proof obligation $(A \land B) \Rightarrow (F \Rightarrow G)$, which is logically equivalent to $(A \land B \land F) \Rightarrow G$.
 
    \item No attached proof, as seen in Figure~\ref{fig:example_input} where the claim \texttt{Main} stands with no proof provided.
\end{itemize}
An important aspect of $\tlaplus$ proof development, which is central to our work, is that appropriate sub-claims must be discovered to establish the parent claim. In Figure~\ref{fig:expected_example_output}, the sub-claims \texttt{FactorForm} and \texttt{SolveFactors} were not given in the original theorem statement (Figure~\ref{fig:example_input}). They had to be formulated by the user with knowledge of quadratic equations. This discovery of effective intermediate steps represents a significant challenge in proof development. Human users must manually determine these sub-claims through mathematical insight and domain knowledge. Our system, however, attempts to automatically discover appropriate sub-claims.

$\tlaplus$ provides several {\em proof directives} that instruct backend provers on how to prove the claims. These include \texttt{OBVIOUS} (indicating that the backend provers should verify the claim directly, as seen in line 9 of Figure~\ref{fig:expected_example_output}), \texttt{BY} (which proves the claim using specified facts and definitions, as demonstrated in line 20), \texttt{BY SMT} (restricting verification to only SMT solvers), and so on. These directives serve as an interface between the high-level proof structure and the specialized reasoning capabilities of various backend provers. 

The status of a claim---whether it is considered \textit{proved} or \textit{unproved}---follows a recursive definition that reflects the hierarchical nature of $\tlaplus$ proofs:
\begin{itemize}[leftmargin=*]
    \item A claim is \textit{proved} if either:
    \begin{itemize}
        \item It has an attached auto proof that is accepted by the backend provers, or
        \item It is supported by a set of sub-claims that are themselves all {\it proved}, and the backend provers confirm that these sub-claims collectively establish the parent claim.
    \end{itemize}
    \item Any claim not meeting these criteria remains \textit{unproved}.
\end{itemize}
This hierarchical structure naturally gives rise to a tree representation of proofs, as visualized in Figure~\ref{fig:proof_tree_example} for the proof shown in Figure~\ref{fig:expected_example_output}. In this tree:
\begin{itemize}[leftmargin=*]
    \item Each node corresponds to a claim (\texttt{Main}, \texttt{FactorForm}, and \texttt{SolveFactors} in our example).
    \item The root node represents the primary goal (\texttt{Main} in our example).
    \item The edges capture the logical dependencies between claims, showing how sub-claims support their parent claims.
\end{itemize}

A proof achieves the status of a \textit{complete proof} precisely when its root goal is \textit{proved} according to the recursive definition above. This completion signifies that the entire proof has been successfully checked by backend solvers.

As constructing formal proofs manually requires significant expertise in both the problem domain and the formal prover itself, there exists a substantial barrier to the wider adoption of formal proving. 
Building on the framework outlined above, the central challenge addressed in this paper is the automated generation of {complete proofs} for $\tlaplus$ {proof obligations}. 

\paragraph{Problem Statement}
Given a module containing an {\em unproved} claim (as in Figure~\ref{fig:example_input} where \texttt{Main} lacks a proof), our objective is to automatically construct a {\em complete proof} (like the one shown in Figure~\ref{fig:expected_example_output}).

    \section{Language Model Guided Proof Automation}
    \label{sec:proof_generation}
    Automated proof generation for $\tlaplus$ presents unique challenges due to its hierarchical proof structure and rigorous verification requirements. In this section, we first describe the main challenges that naive methods face. We then introduce our approach, which leverages the reasoning capabilities of Large Language Models (LLMs) while addressing their limitations through a recursive claim decomposition strategy.

\subsection{Challenges of naive methods}
\label{subsec:motivation_challenges}

We consider two ``naive'' methods:
(1) a symbolic method which simply attempts to use $\TLAPM$ to automatically prove the theorem;
(2) an LLM-based method which prompts the LLM asking for a proof, up to a maximum of $k$ times (this method is actually less naive when $k>1$, as it uses feedback in subsequent prompts).
We discuss each of these two naive methods next.

\subsubsection{Naive symbolic method: $\TLAPM$ \texttt{OBVIOUS}}

The basic approach to automatically proving $\tlaplus$ claims is to delegate them directly to $\TLAPM$'s backend provers without providing further information. In the $\tlaplus$ proof language, this is done by asserting the claim as \texttt{OBVIOUS}. While this method works for simple claims, it often fails for more complex ones that require intermediate proof steps to be explicitly provided.

\begin{algorithm}[t]
    \caption{Direct LLM-Based Proof Generation}
    \label{alg:direct_llm}
    \begin{algorithmic}[1]
        \Function{DirectLLM-ProveObligation}{$\wordcontext, \wordgoal$}
            \State {\em feedback} $\gets$ null
            \Repeat
                \State {\em proof} $ \gets$ \Call{LLMGenProof}{$\wordcontext, \wordgoal, \textit{feedback}$}
                \State {\em proved}, {\em feedback} $ \gets$ \Call{VerifyByTLAPS}{\textit{proof}}
            \Until{{\em proved} or max retries reached}
            \State \Return{{\em proved}, {\em proof}}
        \EndFunction
    \end{algorithmic}
\end{algorithm}

\subsubsection{Direct LLM-Based Proof Generation}
\label{subsec:direct_llm_generation}
A straightforward approach to automated $\tlaplus$ proof generation involves prompting LLMs to generate complete proofs in a single prompt. This method relies entirely on the LLM's ability to produce syntactically correct and logically sound proofs from the provided theorem statements and context.

Algorithm~\ref{alg:direct_llm} outlines this direct approach. For a given proof obligation $(\wordcontext, \wordgoal)$, the algorithm repeatedly prompts the LLM to generate a complete proof (Line 4). After each generation, it verifies the proof using $\TLAPM$ (Line 5). If the proof is valid, the process terminates; otherwise, the algorithm incorporates feedback from the verification step (e.g., error messages) into subsequent prompts to guide the LLM's next attempt (Line 4). This loop continues until a valid proof is found or the maximum number of retries is reached.

Generating $\tlaplus$ proofs directly presents several challenges:

\begin{itemize}[leftmargin=*]
    \item \textbf{Syntactic Correctness:}
    Our experimental results (Section~\ref{subsec:evaluation}) demonstrate that state-of-the-art general-purpose LLMs, including OpenAI o3-mini~\cite{openai_o3_mini} and Google Gemini~\cite{gemini2flash}, frequently generate $\tlaplus$ proofs containing syntax errors, even when provided with the prover's feedback. These errors prevent programmatic verification by $\TLAPM$.
    
    \item \textbf{Monolithic Generation:} 
    When generating complete proofs from a single prompt, LLMs may introduce errors at any point in the proof. Because verification occurs only after the entire proof is generated rather than after each individual step, early errors propagate through subsequent reasoning. This lack of incremental verification limits LLMs' ability to maintain sound reasoning throughout multi-step proofs.
\end{itemize}

Recent approaches such as ReProver~\cite{yang2024leandojo} and COPRA~\cite{thakur2023corpaInContextTheoremProving} address similar challenges in tactic-based theorem provers by constraining LLMs to generate only tactics and premises for a given proof state, enabling step-by-step verification and eliminating syntactic correctness issues.  
However, as discussed in Section~\ref{sec:prelim_problem_statement}, the proof structure and methodology of $\tlaplus$ differ fundamentally from tactic-based provers, necessitating a specialized approach aligned with $\tlaplus$ proof methodology.
Recent studies~\cite{xin2024deepseek_prover_v1_5, wang2024theoremllama} have demonstrated promising results with single-pass Lean proof generation by fine-tuning LLMs on extensive Lean proof corpora, but again are not applicable to $\tlaplus$.
In this paper, we propose a {\em hierarchical proof generation} approach tailored to $\tlaplus$'s proof methodology.

\subsection{System Architecture and Key Ideas}
\label{subsec:sys_architecture}

Our Language Model Guided Proof Automation system (LMGPA) leverages the complementary strengths of LLMs and symbolic methods: we use LLMs for their reasoning abilities to decompose complex claims into simpler sub-claims, while relying on symbolic provers for rigorous verification and for proving simple claims. The key components include:

\begin{itemize}[leftmargin=*]
    \item \textbf{Claim Decomposition:} LLMs guide the decomposition of complex goals into simpler, more manageable sub-claims.
    
    \item \textbf{Automated Proof Generation:} For sufficiently simple claims, the system attempts to generate {\em auto proofs} using $\tlaplus$ directives (e.g., \texttt{OBVIOUS}) that can be directly verified by $\TLAPM$. 
    
    \item \textbf{Proof Validation:} The system uses $\TLAPM$ to verify that (1) sub-claims collectively establish their parent claim, and (2) auto proofs are valid.
\end{itemize}

Our hierarchical, recursive proof generation algorithm (detailed in Section~\ref{subsec:algorithm}) directly addresses the two challenges identified above. First, it mitigates \textit{syntactic correctness} issues by (1) restricting LLMs to generating only claim decompositions rather than complete proofs, and (2) normalizing LLM-generated sub-claim structures (Section~\ref{subsec:llm_claim_decomposition}), which significantly reduces opportunities for syntax errors (Section~\ref{subsec:evaluation}). Second, it overcomes \textit{monolithic generation} limitations through incremental verification at each recursive step, enabling localized error correction without discarding entire proof attempts.

\subsection{Recursive Proof Generation Algorithm}
\label{subsec:algorithm}

Algorithm~\ref{alg:proof_generation} presents our hierarchical proof generation approach. The algorithm recursively decomposes complex claims until reaching claims that can be directly verified by the backend provers, mirroring the hierarchical structure of $\tlaplus$ proofs described in Section~\ref{sec:prelim_problem_statement}.

For a given proof obligation $(\wordcontext, \wordgoal)$, the algorithm first attempts to generate an auto proof (Line 2). If this proof is successfully verified (Line 3), the algorithm returns it immediately. Otherwise, it leverages LLMs to decompose the goal into simpler sub-claims (Line 7) and verifies that these sub-claims collectively establish the original goal (Line 8). This verification-feedback loop continues until either a valid decomposition is found or the maximum number of retries is reached.

Once a valid decomposition is established, the algorithm recursively applies itself to each sub-claim (Lines 14-20), constructing a hierarchical proof structure consistent with $\tlaplus$ proof conventions. 
If \texttt{ProveObligation} fails for any sub-claim (Lines 16-18), the entire proof attempt fails and the algorithm terminates without returning a valid proof, since all sub-claims must be discharged for the overall proof to be valid.

This recursive approach effectively combining the strengths of both symbolic provers (for rigorous verification) and LLMs (for non-trivial claim decomposition) to automate $\tlaplus$ proof generation.

The final proof structure is assembled in Line 21, creating a complete $\tlaplus$ proof that follows the hierarchical structure, with sub-claims serving as lemmas that collectively establish the parent claim.

\begin{algorithm}[h]
    \caption{Hierarchical Proof Generation}
    \label{alg:proof_generation}
    \begin{algorithmic}[1]
        \Function{ProveObligation}{$\wordcontext, \wordgoal$}
            \State $\terminalProof \gets$ \Call{GenerateAutoProof}{$\wordcontext, \wordgoal$}
            \If{VerifyProof($\terminalProof$)}
                \State \Return $\terminalProof$
            \EndIf
            \Repeat
                \State $\wordsubClaims \gets$ \Call{DecomposeIntoSubclaims}{$\wordcontext, \wordgoal$}
                \State ${\worddecompositionValid} \gets$ \Call{VerifyDecomp}{$\wordcontext, \wordgoal, \wordsubClaims$}
            \Until{$\worddecompositionValid$ or max retries reached}
            \If{not $\worddecompositionValid$}
                \State \Return failure
            \EndIf
            \State $\wordproofs \gets \emptyset$
            \ForAll{$\wordclaim \in \wordsubClaims$}
                \State $\wordproof \gets$ \Call{ProveObligation}{$\wordcontext, \wordclaim$}
                \If{$\wordproof$ is failure}
                    \State \Return failure
                \EndIf
                \State $\wordproofs \gets \wordproofs \cup \{(\wordclaim, \wordproof)\}$
            \EndFor
            \State \Return ConstructHierarchicalProof($\wordgoal, \wordsubClaims, \wordproofs$)
        \EndFunction
    \end{algorithmic}
\end{algorithm}

In the following subsections, we delve deeper into the key components of our system, including LLM-guided claim decomposition, auto proof generation, and verification procedures. While developing this system, we explored various optimization techniques beyond those presented here. 
We focus on methods that demonstrated meaningful improvements in our experimental evaluation, while additional optimizations that did not yield significant benefits are documented
\ifdefined\extendedVersion
in Appendix~\ref{appendix:techniques} for completeness.
\else
in the Appendix of the extended version~\cite{zhou2025LMGPAarxiv}.
\fi

\subsection{LLM-Guided Claim Decomposition}
\label{subsec:llm_claim_decomposition}
The \texttt{DecomposeIntoSubclaims} function forms the core of our approach, utilizing pretrained LLMs such as Claude~\cite{claude} and o3-mini~\cite{openai_o3_mini} to identify appropriate intermediate sub-claims that collectively establish a complex parent claim within the hierarchical proof structure. Notably, we use these models without any fine-tuning, relying instead on specialized prompting strategies to guide their reasoning toward valid claim decompositions.

To effectively leverage these models for claim decomposition and to overcome syntactic correctness challenges, we prompt the LLMs to generate normalized sub-claims that adhere to a specific structure%
\ifdefined\extendedVersion
\ (the complete prompt template is available in Appendix~\ref{appendix:template_prompt_for_decomposition}).
\else
\ (the complete prompt template is available in the Appendix of the extended version~\cite{zhou2025LMGPAarxiv}).
\fi

\paragraph{Normalized Claim Structure} 
We constrain the generated sub-claims to follow a normalized format:
\begin{itemize}[leftmargin=*]
    \item Each LLM-generated sub-claim consists of a structured list containing assumptions (boolean expressions or definition references) and a single goal.
    \item Grammar constraints for expressions are embedded in the prompts, restricting output to ASCII characters and providing a table of acceptable notation.
    \item Our system parses these normalized claims and converts them into valid $\tlaplus$ \texttt{ASSUME-PROVE} statements, eliminating a significant source of syntax errors.
\end{itemize}

\paragraph{Adaptive Feedback Loop} 
Although normalization substantially reduces syntax errors, complete correctness cannot be guaranteed. When $\TLAPM$ verification fails, our system feeds back the verifier's output to the LLM, allowing it to generate improved sub-claims in subsequent attempts.

\subsection{Symbolic Auto Proof Generation}
\label{subsec:trivial_proof_generation}

The \texttt{GenerateAutoProof} function employs a heuristic approach to efficiently handle simple claims without querying LLMs. This function first analyzes the parse tree of the given proof module to identify any definitions in the context that need to be explicitly unfolded in the goal claim. Based on this analysis, it generates appropriate auto proofs. If no definitions need unfolding, it applies the $\tlaplus$ directive \texttt{OBVIOUS}, instructing backend provers to attempt verification directly. 
When the system determines that specific definitions must be unfolded to complete a proof, it generates a proof using the $\tlaplus$ directive \texttt{BY $l_1, l_2, ...$ DEF $d_1, d_2, ...$}, where 
$l_1, l_2, ...$ are the assumptions and $d_1, d_2, ...$ are the definitions to be unfolded, both identified through syntax analysis.\footnote{
We also explored a retrieval augmented~\cite{lewis2020RAG} proof generation strategy but our preliminary results did not show sufficient improvements%
\ifdefined\extendedVersion
\ (c.f. Appendix~\ref{appendix:rag_proof_gen}).
\else
\ (c.f. the Appendix of the extended version~\cite{zhou2025LMGPAarxiv}).
\fi
}

\subsection{Verification Procedures}
\label{subsec:verification_procedures}

Our system includes two key verification procedures that ensure the correctness of both auto proofs and claim decompositions:
\paragraph{Auto Proof Verification}
Function \texttt{VerifyProof} directly invokes $\TLAPM$ to determine whether a generated auto proof (e.g., \texttt{OBVIOUS}) is sufficient to justify a claim. 
\paragraph{Decomposition Verification} 
Function \texttt{VerifyDecomp} validates that a set of sub-claims collectively establishes their parent claim. 
The system constructs a $\tlaplus$ module that includes all sub-claims and the parent claim. $\TLAPM$ then verifies that the sub-claims collectively establish the parent claim.

    \section{Implementation and Evaluation}
    \label{sec:experiments}
    We present the implementation details of our system and evaluate its performance on the benchmark suite described in Section~\ref{sec:proof_gen_benchmark}. 

\subsection{Implementation}
We implemented the LMGPA system in Python 3.12, with components for parsing, verification, and LLM interactions. For syntax-level analysis in the auto proof generation phase (Section~\ref{subsec:trivial_proof_generation}), we utilized the Tree-sitter $\tlaplus$ parser~\cite{TLAPlus_tree_sitter}, which enables efficient analysis of $\tlaplus$ parse trees. LLM interactions are managed through LangChain~\cite{LangChain_2022}, providing a unified interface to different language models.

The core verification pipeline integrates the $\TLAPM$ binary~\cite{tlapm_repo}, with wrapper functions that handle the generation of temporary proof modules, execution of verification commands, and parsing of verification results. 
Our prompt templates include detailed instructions on the normalized claim format, examples of correct decompositions, and specific guidance on $\tlaplus$ syntax constraints.
Prompt templates are provided in
\ifdefined\extendedVersion
Appendix~\ref{appendix:prompt_templates}.
\else
\ the Appendix of the extended version of this paper~\cite{zhou2025LMGPAarxiv}.
\fi
 The source code is available on GitHub~\cite{github_src} and the artifact on Zenodo~\cite{artifacts}.

\subsection{Benchmarks}
\label{sec:proof_gen_benchmark}

To evaluate our LMGPA system, we constructed a benchmark suite of $\tlaplus$ theorems drawn from diverse sources to ensure variety in theorem types. The benchmark suite consists of: (a) 93 mathematical theorems adapted from the \texttt{miniF2F}~\cite{zheng2021minif2f} and \texttt{ProofNet}~\cite{azerbayev2023proofnet} collections; plus (b) 26 inductiveness proofs of candidate inductive invariants of distributed protocols, taken from~\cite{schultz2022plainSynthesisTLAplus} and its code repository~\cite{endive_repo}.
\texttt{miniF2F} and \texttt{ProofNet} are standard benchmarks for evaluating AI-powered formal proof generation~\cite{yang2024leandojo}. As these collections lack $\tlaplus$ formalizations, we manually translated a curated subset of these theorems into $\tlaplus$%
\ifdefined\extendedVersion
\ (see Appendix~\ref{appendix:theorem_translation} for details).
\else
\ (for details, see the Appendix of the extended version of this paper~\cite{zhou2025LMGPAarxiv}).
\fi
Both the benchmark suite and our tool are publicly available, as an artifact~\cite{artifacts} as well as in a GitHub repository~\cite{github_src}.

We acknowledge the potential for data leakage in our benchmark. The mathematical theorems from miniF2F and ProofNet are commonly used benchmarks, and while our $\tlaplus$ translations are new, the underlying proof strategies may have been encountered by the LLMs during pretraining in other formal languages such as Lean. Similarly, the inductiveness proofs from~\cite{schultz2022plainSynthesisTLAplus} are publicly available and could appear in pretraining corpora. 
Addressing  potential data leakage issues can be done by constructing a leakage-free $\tlaplus$ proof benchmark suite; this is beyond the scope of the current paper and part of future work.

\subsection{Experimental Setup}
\label{sec:exp_setup}
We evaluated LMGPA on the benchmarks of Section~\ref{sec:proof_gen_benchmark}. For our experiments, we selected state-of-the-art LLMs: 
Claude 3.7 Sonnet~\cite{claude}, Deepseek-V3.2-Exp~\cite{deepseekv3-2-exp}, 
Gemini 2.0 Flash~\cite{gemini2flash}, Gemini 2.5 Flash~\cite{gemini2-5flash}, 
o3-mini-high~\cite{openai_o3_mini}, and GPT-5~\cite{gpt_5}. This selection provides a diverse range of models, including both general language models (Claude and Gemini) and models optimized for reasoning tasks (o3-mini-high), as well as both open-source and proprietary models.
We used all language models without any fine-tuning or additional training, relying solely on prompting strategies to guide these pretrained models.

For all experiments, we set consistent parameters across all models. We limited each model to a maximum of 4 decomposition attempts per claim (Algorithm~\ref{alg:proof_generation}, Line 9) and a maximum of 4 retries per proof obligation for direct LLM proof generation (Algorithm~\ref{alg:direct_llm}, Line 6). To ensure fair comparison and obtain results that are as deterministic as possible, we set the temperature to 0 for all LLM calls. All LLM requests were sent to the API provided by the LLM providers.

All experiments ran on a computer with a 16-core CPU and 64 GB RAM. We configured $\TLAPM$ to use 16 worker processes to fully utilize the available CPU capacity. 
We adhere to the default $\TLAPM$ timeouts. For each proof obligation, $\TLAPM$ attempts three backend provers in sequence: Z3 (5s), Zenon (10s), and Isabelle (30s), resulting in a maximum total timeout of 45s per obligation~\cite{tlaps_online_doc}.

We evaluated our LMGPA system against the following baselines:
\begin{itemize}
    \item Naive symbolic method ($\TLAPM$ \texttt{OBVIOUS}): see Section~\ref{subsec:motivation_challenges}.
    \item Symbolic Auto Proof Generation (SAPG): see Section~\ref{subsec:trivial_proof_generation}.
    \item Direct LLM Proof Generation (DLPG): see Section~\ref{subsec:direct_llm_generation}.
\end{itemize}

\subsection{Results}
\label{subsec:evaluation}

\begin{table}[th]
    \centering
    \caption{Evaluation results on the distributed protocol and mathematical benchmarks (c.f. Section~\ref{sec:proof_gen_benchmark}).
    \textbf{Proved} is the percentage of theorems proved.  \textbf{\#Q} is the total number of queries made to the LLM. \textbf{Time} is the total execution time.
    The best result in each category is highlighted in \textbf{bold}.}
    \label{tab:combined_evaluation_all}
    \setlength{\tabcolsep}{4pt} 
    \begin{tabular}{lcccccc}
        \toprule
        & \multicolumn{3}{c}{\textbf{Distributed protocols}} & \multicolumn{3}{c}{\textbf{Mathematical}} \\
        \cmidrule(lr){2-4} \cmidrule(lr){5-7}
        \textbf{Method} & \textbf{Proved} & \textbf{\#Q} & \textbf{Time} & \textbf{Proved} & \textbf{\#Q} & \textbf{Time} \\
        \midrule
        \texttt{TLAPS OBVIOUS}  & 0.0\% & none & 1m & 44.1\% & none & 25m \\
        SAPG                    & 38.5\% & none & 14m & 49.5\% & none & 34m \\
        \midrule
        DLPG[Claude-3.7-Sonnet] & \textbf{15.4\%} & 98 & 4h 43m & 17.2\% & 321 & 5h 14m \\
        DLPG[Deepseek-V3.2-Exp] & 0.0\% & 104 & 11h 27m & \textbf{29.0\%} & 336 & 39h 51m \\
        DLPG[Gemini-2.0-Flash]  & 0.0\% & 104 & 41m & 4.3\% & 359 & 39m \\
        DLPG[Gemini-2.5-Flash]  & 0.0\% & 104 & 1h 30m & 3.2\% & 364 & 3h 18m \\
        DLPG[GPT-5]             & 3.8\% & 104 & 6h 36m & 20.7\% & 307 & 24h 23m \\
        DLPG[o3-mini-high]      & 0.0\% & 104 & 1h 5m & 0.0\% & 372 & 8h 30m \\
        \midrule
        LMGPA[Claude-3.7-Sonnet]    & 42.3\% & 83 & 1h 32m & 53.8\% & 231 & 4h 49m \\
        LMGPA[Deepseek-V3.2-Exp]    & \textbf{50.0\%} & 73 & 9h 14m & \textbf{59.1\%} & 261 & 33h 17m \\
        LMGPA[Gemini-2.0-Flash]     & 38.5\% & 54 & 39m & 54.8\% & 251 & 1h 10m \\
        LMGPA[Gemini-2.5-Flash]     & 46.2\% & 72 & 1h 25m & 54.8\% & 224 & 5h 2m \\
        LMGPA[GPT-5]                & 42.3\% & 89 & 4h 26m & 58.1\% & 245 & 9h 58m \\
        LMGPA[o3-mini-high]         & 42.3\% & 96 & 1h 44m & 57.0\% & 241 & 5h 50m \\
        \bottomrule
    \end{tabular}
\end{table}

Our evaluation focuses on three metrics: (1) the percentage of theorems successfully proved, which is our primary effectiveness measure, (2) the total number of LLM queries, and (3) the total time taken to process the entire benchmark suite.

The results are shown in Table~\ref{tab:combined_evaluation_all}. 
Across all benchmarks and all of the tested LLMs, our LMGPA system demonstrates consistent improvements in proof success rates compared to the baselines. The SAPG baseline itself consistently outperforms the \texttt{OBVIOUS}-only baseline, demonstrating the effectiveness of our heuristic-based symbolic proof generation component.

\paragraph{Timing Considerations}
The timing differences are heavily influenced by factors unrelated to the models' capabilities. Network conditions, model architecture, caching strategies, and the hardware infrastructure of different model providers all significantly impact execution times---making raw timing comparisons between models less meaningful for evaluating proof generation effectiveness. The total time for evaluating the entire benchmark suite is thus reported for completeness.

\paragraph{Comparison with Combined Baselines}

To further understand our approach, we compared it against a combined baseline that represents the best performance achievable by either baseline independently: see Table~\ref{tab:autoproof_plus_directllm_eval_comp}. Specifically, we consider a theorem as proved by the combined {\it SAPG+DLPG} baseline if either the SAPG or the DLPG successfully proves it.
We also define a {\em total combined} approach, which considers a theorem as proved if it is successfully proved by any of the three methods, SAPG, or DLPG, or LMGPA.

\begin{table}[ht]
    \centering
    \caption{Comparison between combined baseline and our system}
    \label{tab:autoproof_plus_directllm_eval_comp}
    \begin{tabular}{lccc}
        \toprule
        \multirow{2}{*}{\textbf{Model}} & \multicolumn{3}{c}{\textbf{Proved}} \\
        \cmidrule(lr){2-4}
        & {\it SAPG+DLPG~} & {\it LMGPA~} & {\it {Total Combined}} \\
        \midrule
        Claude-3.7-Sonnet    &  52.9\% & {51.3\%}    & 54.6\%      \\
        {Deepseek-V3.2-Exp}  &  52.9\% & {57.1\%}    & 61.3\%      \\
        Gemini-2.0-Flash     &  49.6\% & {51.3\%}    & 52.9\%      \\
        {Gemini-2.5-Flash}   &  49.6\% & {52.9\%}    & 55.5\%      \\
        {GPT-5}              &  54.6\% & {54.6\%}    & 62.2\%      \\
        o3-mini-high         &  47.1\% & {53.8\%}    & 53.8\%      \\
        \bottomrule
    \end{tabular}
\end{table}

\paragraph{Syntax Errors in LLM-Generated Content}

We also analyzed the syntactic validity of the content generated by LLMs in both DLPG and LMGPA systems. Because the generation targets differ, we aligned our evaluation with the specific output of each LLM query. For DLPG, we evaluated the full proofs, whereas for LMGPA, we evaluated the decompositions (sub-claims). We focused on decompositions for LMGPA because other proof structures are generated symbolically; thus, checking the decomposition isolates the LLM's actual contribution. Table~\ref{tab:syntactic_correctness_comparison} shows that while DLPG suffers from low syntactic validity, LMGPA achieves significantly higher syntactic validity rates.
These results suggest that most of DLPG's failures stem from syntax errors, which aligns with our motivation for using normalized claim structure.

\begin{table}[tbh]
    \centering
    \caption{Comparison of Syntactic Validity of LLM-Generated Proofs Between Approaches}
    \label{tab:syntactic_correctness_comparison}
    \scalebox{0.85}{
    \begin{tabular}{lcccc}
        \toprule
        \multirow{2}{*}{\textbf{Model}} & \multicolumn{2}{c}{\textit{DLPG}} & \multicolumn{2}{c}{\textit{LMGPA}} \\
        \cmidrule(lr){2-3} \cmidrule(lr){4-5}
        & Syn. Valid/\#Queries & ~Percentage~ & Syn. Valid/\#Queries & ~Percentage~ \\
        \midrule
        Claude-3.7-Sonnet         & 88/434 & 20.3\% & 206/314 & 65.6\% \\
        Deepseek-V3.2-Exp         & 84/425 & 19.8\% & 287/334 & 85.9\% \\
        Gemini-2.0-Flash          & 27/463 &  5.8\% & 195/305 & 63.9\% \\
        Gemini-2.5-Flash          & 4/468  &  0.9\% & 228/296 & 77.0\% \\
        GPT-5                     & 66/411 & 16.1\% & 290/334 & 86.8\% \\
        o3-mini-high              & 1/476  &  0.2\% & 252/337 & 74.8\% \\
        \bottomrule
    \end{tabular}
    }
\end{table}

\paragraph{Failures Due to Prover Limitations} 

\begin{figure}[ht]
    \centering
        \begin{minipage}[t]{0.96\textwidth}
            \centering
            \setlength{\abovecaptionskip}{0pt}
            \begin{lstlisting}[frame=single,style=tlaplusfancy]
---- MODULE exercise_1_27 ----
EXTENDS Integers, TLAPS

Cube(x) == x * x * x

THEOREM L1 == \E x, y, z, w \in Int : Cube(x) + Cube(y) = 1729 /\ Cube(z) + Cube(w) = 1729 /\ x # z /\ x # w /\ y # z /\ y # w
THEOREM L2 == \A n \in Nat : (n < 1729) => ~(\E x, y, z, w \in Int : Cube(x) + Cube(y) = n /\ Cube(z) + Cube(w) = n /\ x # z /\ x # w /\ y # z /\ y # w)
THEOREM exercise_18_4 == \A n \in Nat : (\E x, y, z, w \in Int : Cube(x) + Cube(y) = n 
      /\ Cube(z) + Cube(w) = n /\ x # z /\ x # w /\ y # z /\ y # w)  => n >= 1729
BY L1, L2 DEF Cube
======================\end{lstlisting}
    \caption{An example of $\TLAPM$ limitation. Theorem \texttt{L2} is the contrapositive of the target theorem, making the implication ($\texttt{L1} \land \texttt{L2} \Rightarrow \texttt{exercise\_18\_4}$) trivially valid. However, $\TLAPM$ fails to prove this within the timeout.}
    \label{fig:good_decomposition_but_cannot_verify}
    \end{minipage}
\end{figure}

Beyond syntax errors and logically incorrect decompositions, another failure mode in LMGPA arises
when LLMs generate mathematically valid decompositions that $\TLAPM$ fails to verify automatically.
Figure~\ref{fig:good_decomposition_but_cannot_verify} shows an LLM-generated decomposition attempt of theorem \texttt{exercise\_18\_4} into sub-claims \texttt{L1} and \texttt{L2}. Observe that \texttt{L2} is the contrapositive of the target theorem. Thus, the decomposition is logically valid: $(\texttt{L1} \land \texttt{L2}) \Rightarrow \texttt{exercise\_18\_4}$ holds (the assumption \texttt{L1} is redundant, but this does not affect the validity of the implication). However, $\TLAPM$ cannot prove this implication within the default timeout.

\paragraph{Quality of Generated Proofs}

\begin{figure}[ht]
    \centering
        \begin{minipage}[t]{0.96\textwidth}
            \centering
            \setlength{\abovecaptionskip}{0pt}
            \begin{lstlisting}[frame=single,style=tlaplusfancy]
---- MODULE mathd_numbertheory_234 ----
EXTENDS TLAPS, Integers

THEOREM CubeOf97_1 == 97 * 97 * 97 = 912673
OBVIOUS

THEOREM OnlySolution_2 == 
ASSUME NEW a \in Nat, NEW b \in Nat, a >= 1, a <= 9, b <= 9, 
    (10 * a + b) * (10 * a + b) * (10 * a + b) = 912673
PROVE 10 * a + b = 97
OBVIOUS

THEOREM UniqueDigits_3 == 
ASSUME NEW a \in Nat, NEW b \in Nat, a >= 1, a <= 9, b <= 9, 10 * a + b = 97
PROVE a = 9 /\ b = 7
OBVIOUS

THEOREM SumIs16_4 == 
ASSUME NEW a \in Nat, NEW b \in Nat, a = 9, b = 7 
PROVE a + b = 16
OBVIOUS

THEOREM mathd_numbertheory_234 == 
\A a, b \in Nat :
        (a >= 1) /\ (a <= 9) /\ (b <= 9) /\ 
        ((10 * a + b) * (10 * a + b) * (10 * a + b) = 912673) 
        => (a + b = 16)
BY CubeOf97_1, OnlySolution_2, UniqueDigits_3, SumIs16_4
==================================\end{lstlisting}
    \caption{A valid but verbose proof generated by LMGPA: sub-claims \texttt{CubeOf97\_1} and \texttt{SumIs16\_4} are both trivial and redundant.}
    \label{fig:verbose_proof_example}
    \end{minipage}
\end{figure}

We  qualitatively analyzed the proofs generated by LMGPA. The generated proofs are generally well-structured, following $\tlaplus$ syntax with hierarchical organization of claims and sub-claims (e.g., Figure~\ref{fig:expected_example_output} and~\ref{fig:verbose_proof_example}).
However, they do not yet match the quality of human-written proofs.
Some proofs, while logically valid, do not correspond to the most concise and natural way to prove the theorem.
For example, in the proof shown in Figure~\ref{fig:verbose_proof_example}, sub-claims \texttt{CubeOf97\_1} and \texttt{SumIs16\_4} are trivial calculations that are unnecessary in establishing the goal.
We observed similar patterns elsewhere: decompositions that introduce more sub-claims than necessary or use intermediate claims uncommon in human proofs.
Also, in some cases decompositions are not useful. For example, in Figure~\ref{fig:good_decomposition_but_cannot_verify}, \texttt{L2} is a restatement of the goal, so the LLM-generated decomposition does not meaningfully simplify the target theorem.
These observations suggest room for improvement in generating more concise and natural proofs, as well as better decompositions.
We nevertheless note that our system is capable of generating non-trivial proofs.
For example, a distributed protocol inductiveness proof generated by our LMGPA system is 227 lines long (including 88 lines of definitions).
\ifdefined\extendedVersion
This entire proof is listed in Appendix~\ref{appendix:example_proofs_found}.
\else
\ This entire proof is listed in the Appendix of the extended version of this paper~\cite{zhou2025LMGPAarxiv}.
\fi
All proofs generated in our experiments can be found in the public repository~\cite{github_src} as well as in the artifact of this paper~\cite{artifacts}.

    \section{Related Work}
    \label{sec:related_work}
    
\paragraph{LLM-assisted Theorem Proving}
Recent years have seen significant advances in applying LLMs to formal reasoning tasks. In the domain of theorem proving,
GPT-f~\cite{polu2020gpt_f} is a generative language model for automated theorem proving using Metamath~\cite{megill2019metamath}.
Baldur~\cite{first2023baldur} shows the LLMs' abilities on generating and repairing formal proofs in the Isabelle/HOL.
Tahat et al.~\cite{tahat2024proof_repair_case_study} present a case study on proof repair utilizing LLMs in Coq.
LeanDojo~\cite{yang2024leandojo} demonstrates the use of language models and retrieval-augmented generation for generating proof tactics and selecting premises in the Lean theorem prover, providing both tactical suggestions and a comprehensive benchmark suite for evaluating LLMs on formal proof tasks. 
COPRA~\cite{thakur2023corpaInContextTheoremProving} applies in-context learning to both Rocq and Lean provers, demonstrating how learning from existing examples can improve proof generation in these provers.
Hilbert~\cite{varambally2025hilbert} leverages this idea and uses both general purpose and specialized LLMs for different levels of proof generation in Lean.
Cobblestone~\cite{kasibatla2024cobblestone} presents a divide-and-conquer approach for synthesizing Rocq proofs using off-the-shelf LLMs.
Unlike these approaches, our method targets generating $\tlaplus$ proofs.

Liang et al.~\cite{liang2025gLLM_decomp} argue
that general purpose LLMs perform well on high-level proof decomposition comparing to specialized models fine-tuned for theorem proving tasks.
Zhang et al.~\cite{zhang2023getting_more_out_of_llm_for_proofs} give
a detailed analysis of LLMs' capabilities in formal theorem proving and proposes general suggestions to enhance their performance.
Despite the successes in LLM-assisted reasoning, Mirzadeh et al.~\cite{mirzadeh2024gsm} discuss the limitations of LLMs in mathematical reasoning.

Fine-tuning approaches have also shown promise, with DeepSeek-Prover-V1.5~\cite{xin2024deepseek_prover_v1_5} and TheoremLlama~\cite{wang2024theoremllama} achieving notable results in single-pass Lean proof generation through specialized training on extensive Lean proof corpora. 
POETRY~\cite{wang2024POETRY} generates Isabelle proofs recursively using a fine-tuned language model, whereas our approach leverages general-purpose LLMs to generate $\tlaplus$ proofs.
Rango~\cite{thompson2025rango} fine-tunes a language model to generate Rocq tactics step by step, augmenting the generation with retrieved similar proofs and relevant lemmas.
Other approaches include LEGO-Prover~\cite{wang2023lego_prover}, which employs a growing library of verified lemmas to augment LLMs' theorem proving capabilities, and work by Jiang et al.~\cite{jiang2022informal_proof}, which maps informal proofs to formal proof sketches that guide automated provers.
DeepSeek-Prover-V2~\cite{ren2025deepseek-prover-v2} explores subgoal decomposition strategies via reinforcement learning to enhance formal reasoning capabilities for LLMs in Lean. 
Our work differs from these approaches by focusing on $\tlaplus$, which employs a different proof structure.

These advances have been supported by standardized benchmarks such as miniF2F~\cite{zheng2021minif2f} and ProofNet~\cite{azerbayev2023proofnet}, which provide diverse collections of mathematical problems for evaluating theorem provers across different formal systems. 

\paragraph{LLMs for Software Verification}
Beyond mathematical theorem proving, LLMs have shown promise in software verification tasks. Clover~\cite{sun2023clover} leverages LLMs to generate Dafny code and annotations, while Chakraborty et al.~\cite{chakraborty2023rankingLoopInvariantsForVerification} apply them to loop invariant generation.
Wen et al.~\cite{wen2024enchantingProgramSpecificationSynthesis} combines LLMs with static analysis tools for program specification synthesis, and the Lemur system~\cite{wu2023lemur} demonstrates how LLMs can enhance traditional program verification frameworks.
Laurel~\cite{mugnier2025laurel} provides a framework for using LLMs to generate and verify program specifications in Dafny and a benchmark extracted from real-world codebase.
DafnyBench~\cite{loughridge2024dafnybench} provides a benchmark suite for evaluating LLMs in the context of Dafny program verification.
Selene~\cite{zhang2024selene} proposes a benchmark for automated software verification, grounded in seL4 kernel~\cite{klein2009sel4}.

\paragraph{Prompt Engineering and In-Context Learning}
Research has explored LLMs' capabilities in general reasoning tasks~\cite{ho2022large, zhang2024llm} and the role of prompt engineering in formal methods applications~\cite{chen2023nl2tl,cosler2023nl2spec}. Techniques such as in-context learning~\cite{dong2022survey_in_context_learning, rubin2021learning} and dynamic prompt adjustment~\cite{wei2022CoT,yao2024ToT} have proven effective in improving LLMs' performance on tasks requiring precise logical reasoning.

    \section{Conclusion}
    \label{sec:conclusion}

    We present a language model-guided approach for automating $\tlaplus$ proof generation through hierarchical decomposition of complex proof obligations. Our key insight is that by constraining LLMs to generate normalized claim decompositions rather than complete proofs, we can leverage their reasoning capabilities while mitigating their tendency to produce syntactically incorrect formal proofs. Our evaluation shows substantial gains over direct LLM proof generation while highlighting the importance of combined LLM+symbolic tools.

    Future work includes exploring specialized training methods, such as fine-tuning on $\tlaplus$ proof corpora, to address persistent syntax errors, and to improve decomposition quality and the overall success rates. We also plan to investigate advanced prompting strategies and retrieval-augmented techniques. 
    Another direction for future work is investigating how to guide LLMs to generate decompositions that are not only mathematically valid but also readily verifiable by automated provers. 
    Training or guiding LLMs to understand the capabilities and limitations of symbolic provers could lead to more effective proof automation strategies.

    \section*{Acknowledgments}
    We would like to thank William Schultz and Ian Dardik who provided the inductive invariant proofs that formed our distributed protocol benchmarks.
        This work has been supported by NSF's FMitF (Formal Methods in the Field) program, under awards 2319500 and 2525087.

    \bibliographystyle{splncs04}
    \bibliography{ref}

    \ifdefined\extendedVersion
        \newpage
        \appendix
        \section*{Appendix}
        \section{Other Techniques Used}
\label{appendix:techniques}

\subsection{Proof Context Management Optimization}
\label{appendix:proof_context_optimization}

An optimization we tried is the management of proof context to reduce reasoning complexity and verification time. As discussed in Section~\ref{sec:prelim_problem_statement}, $\TLAPM$ requires explicit references to facts and definitions used in proofs rather than automatically considering all available information.

Before the recursive calls to \texttt{ProveObligation} function (line 15 in Algorithm~\ref{alg:proof_generation}), we minimize the proof context by extracting only the necessary definitions and facts from the full proof context. Specifically, we:
\begin{itemize}[leftmargin=*]
    \item Extract all operators, functions, and constants referenced in the claim
    \item Identify their definitions in the context
\end{itemize}
This reduced context is used both in prompt construction for LLMs and in verification calls to $\TLAPM$, resulting in:
\begin{itemize}[leftmargin=*]
    \item Shorter, more focused prompts that help LLMs concentrate on relevant information
    \item Improved performance of backend provers by eliminating unnecessary context
\end{itemize}
This context minimization represents an important practical consideration for deploying our system on real-world proof obligations, where the full context might include numerous definitions and theorems not directly relevant to the specific claim being proved.

However, this optimization does not show significant improvements in our preliminary experiments. We hypothesize that it is because the full context is already relatively small in our benchmark suite. Thus, the benefits of context minimization are not as pronounced as expected. We leave further exploration of this optimization for future work.

\subsection{Retrieval Augmented Auto Proof Generation}
\label{appendix:rag_proof_gen}

\begin{wrapfigure}{R}{0.4\textwidth}
    \centering
    \resizebox{0.95\linewidth}{!}{\begin{tikzpicture}[
    node distance=0.3cm,
    box/.style={rectangle, draw, minimum width=2.5cm, minimum height=0.7cm, font=\footnotesize},
    process/.style={rectangle, fill=orange!70, rounded corners, minimum width=2.5cm, minimum height=0.7cm, font=\footnotesize},
    arrow/.style={-{Stealth[length=1.5mm]}, thick},
    dash box/.style={rectangle, draw=gray, dashed, rounded corners, minimum width=2.5cm, minimum height=0.7cm, font=\footnotesize}
]

\node[font=\small\itshape] (title) {Retrieval-Augmented Auto Proof Generation};

\node[box, below=0.4cm of title] (obl) {Proof Obligation $Obl$};
\node[process, below=0.3cm of obl] (db) {1. Query Proof Database};
\node[dash box, below=0.3cm of db] (top) {top-$k$ similar proofs};
\node[process, below=0.3cm of top] (synth) {2. Construct Prompt};
\node[process, below=0.3cm of synth] (llm) {3. Query LLM};
\node[box, below=0.3cm of llm] (proof) {Auto Proof of $Obl$};

\draw[arrow] (obl) -- (db);
\draw[arrow] (db) -- (top);
\draw[arrow] (top) -- (synth);
\draw[arrow] (synth) -- (llm);
\draw[arrow] (llm) -- (proof);

\draw[arrow] (obl) to[out=-10, in=10] (synth);

\begin{scope}[on background layer]
\node[fit=(db) (top) (synth) (llm), draw, rounded corners, dotted, inner sep=0.15cm] (bg) {};
\end{scope}
\end{tikzpicture}}
    \caption{Workflow of the Retrieval-Augmented Auto Proof Generation approach.}
    \label{fig:apdx_rag_proof_gen}
\end{wrapfigure}
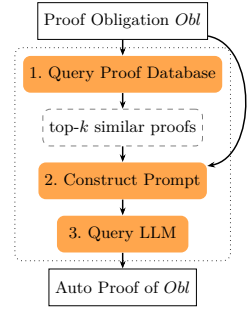

In addition to the approach described in the main paper, we explored a Retrieval-Augmented Generation (RAG) technique to enhance auto proof generation. While this approach did not improve success rates in our preliminary experiments, we document it here for completeness. We will explore this direction further in future work.

\subsubsection{Motivation and Approach}

The Auto Proof Generation (described in Section~\ref{subsec:trivial_proof_generation}) focuses on producing valid auto proofs for simple claims without requiring further decomposition. 
One limitation of our heuristic method is that it always unfolds all available definitions and includes all facts in the context, which might not be the optimal option. A more selective use of only necessary definitions and facts could potentially improve the prover's performance. We hypothesize that a RAG-based approach combined with LLMs might help identify which definitions and facts are truly necessary for a proof, avoiding the use of unnecessary definitions that could complicate the proving process.

To test this hypothesis, we implemented a RAG approach that leverages a database of verified $\tlaplus$ proofs to guide LLM-based proof generation. As illustrated in Figure~\ref{fig:apdx_rag_proof_gen}, our approach consisted of three main steps: (1) retrieving similar proofs from a proof database, (2) synthesizing a prompt with these examples, and (3) generating candidate proofs using an LLM.

\subsubsection{Proof Database Construction}

We constructed a proof database containing proof statements extracted from verified $\tlaplus$ specifications in the $\tlaplus$ Examples repository~\cite{tlaplus_examples}. A \textit{proof statement} refers to the text containing a claim and its proof, typically represented by a theorem or lemma with its corresponding proof directive (e.g., \texttt{OBVIOUS}, \texttt{BY DEF}, etc.).

For example, for the following proof statement in the $\TLAPM$'s standard library:
\begin{lstlisting}[frame=single,style=tlaplusfancy]
THEOREM FS_SameCardinalityBij ==
  ASSUME NEW S, NEW T, IsFiniteSet(S), IsFiniteSet(T),
         Cardinality(S) = Cardinality(T)
  PROVE  ExistsBijection(S,T)
BY FS_CardinalityType, Fun_ExistsBijSymmetric, Fun_ExistsBijTransitive\end{lstlisting}
We will extract the \texttt{ASSUME-PROVE} struct and store the facts used in \texttt{BY} clause to the database. Thus, when we query the database with a similar claim, it is able to retrieve this \texttt{BY} proof for reference.

\subsubsection{Similarity-Based Retrieval}

Given a claim requiring a proof, our retrieval process identified similar proof statements from the database through semantic similarity matching:
\begin{itemize}[leftmargin=*]
    \item We used an embedding function $f$ to map each \texttt{ASSUME-PROVE} struct to a vector in an $n$-dimensional space, computing embeddings $\mathbf{v}_{\textrm{claim}}$ for the target claim and $\mathbf{v}_{i}$ for each statement in the database.
    
    \item We used a pretrained text embedding to compute these vector representations.
    
    \item Using cosine similarity:
    $$
    \textrm{Sim}(\mathbf{v}_{\textrm{claim}}, \mathbf{v}_{i}) = \frac{\mathbf{v}_{\textrm{claim}} \cdot \mathbf{v}_{i}}{\|\mathbf{v}_{\textrm{claim}}\| \|\mathbf{v}_{i}\|},
    $$
    we selected the $k$ proof statements with highest similarity scores to form a {\it reference set}.
\end{itemize}

\subsubsection{RAG-Enhanced Proof Generation}

Using the retrieved reference set, we constructed a prompt that included: (1) The target claim to be proved, (2) The $k$ most similar claims and their proofs, and (3) instructions for generating a valid $\tlaplus$ proof.

We then used this prompt with an LLM to generate candidate proofs. Multiple candidates were generated in parallel to increase the likelihood of finding a valid proof. Each candidate was verified using $\TLAPM$, and the first valid proof was selected.

\subsubsection{Experimental Results}
A key challenge with this approach is the non-deterministic nature of LLMs, which makes it difficult to guarantee syntactic correctness of generated proofs. We observed that irrelevant information from retrieved examples occasionally confused the LLM, resulting in less effective proofs. To address this issue, we implemented a fallback mechanism that defaulted to our heuristic approach described in Section~\ref{subsec:trivial_proof_generation} when the RAG-generated proofs failed verification.

Our preliminary experiments revealed that the heuristic approach alone achieved comparable success rates without the additional complexity of the RAG system. Despite the theoretical advantage of more selective use of definition and fact, this benefit did not show in measurable performance improvements. We hypothesize that a more comprehensive proof benchmark suite and refined retrieval techniques might be necessary for this approach to demonstrate its potential value in future work.

\subsection{Falsification for Sub-claim Validation}
In addition to the verification procedures described in Section~\ref{subsec:verification_procedures}, we implemented a falsification step to enhance the robustness of sub-claim validation during decomposition.
For each generated sub-claim, the system attempts to falsify it by proving its negation. If a sub-claim's negation is proven valid, the decomposition is immediately rejected as the sub-claim would be trivially false.

However, in our experiments on the benchmark suite, this falsification step did not identify any invalid sub-claims beyond those already caught by the existing verification procedures. While the falsification step provides an additional safety check, it did not improve performance on our current benchmarks. This suggests that either the LLM-generated decompositions rarely produce trivially false sub-claims, or that the existing verification procedures are already sufficient to detect problematic decompositions through their failure to collectively establish the parent claim.

\section{Benchmarks}
\subsection{Benchmark Suite Organization and Utilities}
\label{sec_benchmark_organization}

The benchmark suite is structured as a collection of $\tlaplus$ modules, each contained in a separate file with a single unproved theorem. 
For evaluation and analysis, we also provide detailed metadata for each module. This metadata is used only for evaluation and is not an input to our proof automation system. Instead, our system automatically extracts this information, which includes: (1) the goal theorem's name and complete specification, (2) context information encompassing all relevant definitions and lemmas, and (3) the line number where a proof for the goal theorem should be inserted.

To support benchmark suite extensibility, we developed utilities that automate the extraction of theorems and contextual information from $\tlaplus$ files. These tools automatically identify unproved theorems and generate the corresponding metadata, enabling researchers to easily incorporate additional theorems into the benchmark suite.

This benchmark suite enables fair comparison between different proof automation approaches and establishes baseline performance metrics for future $\tlaplus$ proof automation research.

\subsection{Manual vs LLM Translation of Theorems into $\tlaplus$}
\label{appendix:theorem_translation}

The \texttt{miniF2F}~\cite{zheng2021minif2f} and \texttt{ProofNet}~\cite{azerbayev2023proofnet} collections lack $\tlaplus$ formalizations, so we had to translate our benchmarks from these collections into $\tlaplus$. Many original theorems relied on mathematical objects not supported by standard $\TLAPM$ libraries (e.g., real arithmetic/groups) and were not included in our benchmarks. We prioritized theorems involving concepts like factorials and prime numbers, which can be expressed using natural numbers and recursive functions supported by current libraries. This resulted in 93 mathematical theorems (81 from \texttt{miniF2F} and 12 from \texttt{ProofNet}).

We explored using LLMs to automatically translate theorems from other proof assistants to $\tlaplus$ for our benchmark suite construction. While this approach seemed promising for efficiently expanding our benchmark, our experiments revealed significant limitations that led us to adopt manual translation instead.

\begin{figure}[htbp]
    \centering
        \begin{minipage}[t]{0.46\textwidth}
            \centering
            \setlength{\abovecaptionskip}{0pt}
            \begin{lstlisting}[frame=single,style=tlaplusfancy]
---- MODULE exercise_1_27 ----
EXTENDS Integers, TLAPS
(*
Original Lean 4 Theorem:
theorem exercise_1_27 {n : Nat} (hn : Odd n) : 8 | (n^2 - 1) := by
  -- Proof details omitted
*)
(* automatically translated to the TLA+ specification: *)
THEOREM exercise_1_27 ==
  \A n \in Nat : (\E k \in Nat : n = 2*k + 1) => 8 | (n^2 - 1)
====\end{lstlisting}
            \caption{An example of incomplete translation from Lean to $\tlaplus$, generated by an LLM. The generated $\tlaplus$ specification contains the definition of odd numbers but lacks the divisibility relation \texttt{|}.}
            \label{fig:apdx_auto_trans_spec}
        \end{minipage}
    \hfill
    \begin{minipage}[t]{0.46\textwidth}
        \centering
        \setlength{\abovecaptionskip}{0pt}
        \begin{lstlisting}[frame=single,style=tlaplusfancy]
---- MODULE exercise_1_27 ----
EXTENDS Naturals, TLAPS
(*
Original Lean 4 Theorem:
theorem exercise_1_27 {n : Nat} (hn : Odd n) : 8 | (n^2 - 1) := by
  -- Proof details omitted
*)

Odd(n) == n % 2 = 1
Divides(a, b) == \E k \in Nat : b = a * k

THEOREM exercise_1_27 ==
  \A n \in Nat : 
    Odd(n) => Divides(8, (n * n - 1))
====\end{lstlisting}
        \caption{Manual translation of the Lean theorem in Figure~\ref{fig:apdx_auto_trans_spec} into $\tlaplus$.}
        \label{fig:apdx_manual_trans_spec}
    \end{minipage}
\end{figure}

The LLM-based translation attempts consistently produced specifications with various deficiencies, as illustrated in Figure~\ref{fig:apdx_auto_trans_spec}. Common issues included:

\begin{itemize}[leftmargin=*]
    \item \textbf{Incomplete translations}: Many generated specifications omitted necessary definitions or used undefined symbols, as shown by the missing definition of the divisibility relation ("\texttt{|}") in Figure~\ref{fig:apdx_auto_trans_spec}.
    
    \item \textbf{Syntax errors}: LLMs frequently produced $\tlaplus$ code with invalid syntax that could not be parsed by $\TLAPM$, as shown in Figure~\ref{fig:apdx_incorrect_auto_trans_spec}, which contains invalid operators ("\texttt{!}") and unsupported Unicode.
\end{itemize}

Moreover, we found no straightforward way to automatically verify the correctness of these translations. Determining whether a translation preserves the original theorem's mathematical meaning would still require manual inspection.

Given these challenges, manual translation (as shown in Figure~\ref{fig:apdx_manual_trans_spec}) proved to be the most reliable approach for creating our benchmark. This decision prioritized quality and correctness over quantity, ensuring that our benchmark suite contains valid $\tlaplus$ specifications that accurately represent the original mathematical problems.

\begin{figure}[t]
    \centering
    \begin{minipage}[t]{0.46\textwidth}
        \centering
        \setlength{\abovecaptionskip}{0pt}
        \begin{lstlisting}[frame=single,style=tlaplusfancy]
---- MODULE MathdNumbertheory559 ----
EXTENDS TLAPS, Integers, FiniteSets

(*
Original Lean 4 theorem:
theorem mathd_numbertheory_559 (x y : <@$\mathbb{N}$@>) (h<@$_0$@> : x % 3 = 2) (h<@$_1$@> : y % 5 = 4) (h<@$_2$@> : x % 10 = y % 10) :
    14 <@$\le$@> x := by
*)

VARIABLES x, y

ASSUME h<@$_0$@> == x % 3 = 2
ASSUME h<@$_1$@> == y % 5 = 4
ASSUME h<@$_2$@> == x % 10 = y % 10

THEOREM MathdNumbertheory559 ==
    ASSUME NEW x <@$\in$@> Nat, NEW y <@$\in$@> Nat,
           h<@$_0$@>! (x % 3 = 2),
           h<@$_1$@>! (y % 5 = 4),
           h<@$_2$@>! (x % 10 = y % 10)
    PROVE  14 <@$\le$@> x
    OBVIOUS
====\end{lstlisting}
        \caption{An example of syntactically incorrect $\tlaplus$ code generated by an LLM. The Unicode identifiers like $h_0$ are not natively supported in $\TLAPM$. The use of operator \texttt{!} in line 18 is syntactically invalid.}
        \label{fig:apdx_incorrect_auto_trans_spec}
    \end{minipage}
    \hfill
    \begin{minipage}[t]{0.46\textwidth}
        \centering
        \setlength{\abovecaptionskip}{0pt}
        \begin{lstlisting}[frame=single,style=tlaplusfancy]
---- MODULE amc12a_2002_p6 ----
EXTENDS Integers

THEOREM amc12a_2002_p6 ==
  \A n \in Nat \ {0} :
    \E m \in Nat :
      (m > n) /\ (\E p \in Nat : m * p <= m + p)
PROOF
<1>1. FIX n \in Nat \ {0}.
<1>2. TAKE m = n + 1.
<1>3. HAVE m > n BY INT_ARITH.
<1>4. TAKE p = 1.
<1>5. HAVE m * p <= m + p BY INT_ARITH.
<1>6. QED
====\end{lstlisting}
        \caption{An example of a syntactically incorrect $\tlaplus$ proof generated by o3-mini-high. The proof includes the \texttt{FIX} construct, which {is valid in Isabelle/Isar but} undefined in $\tlaplus${, indicating confusion between proof assistant syntaxes.} The periods at the end of each proof line are also invalid in $\tlaplus$ syntax.}
        \label{fig:apdx_syntactically_incorrect_proof}
    \end{minipage}
\end{figure}

\section{Syntactically Incorrect Proofs Generated by LLMs}
\label{appendix:syntactically_incorrect_proof}

Figure~\ref{fig:apdx_syntactically_incorrect_proof} demonstrates a typical example of syntax errors in proofs generated by LLMs when evaluating the direct LLM proof generation baseline. This specific example was generated by o3-mini-high when tasked with proving a theorem about natural numbers.

The generated proof contains several syntactic constructs that are incompatible with $\TLAPM$. The proof uses the keyword \texttt{FIX} (line 9) which does not exist in $\TLAPM$'s proof language. Additionally, each proof step incorrectly ends with a period, which is not valid $\TLAPM$ syntax and causes parsing errors. Furthermore, the proof attempts to use \texttt{INT\_ARITH} as a proof strategy (lines 10 and 11), which suggests confusion with other proof assistants' automated tactics.

This example illustrates one of the primary challenges discussed in Section~\ref{subsec:evaluation}: LLMs frequently mix syntax from different proof assistants when attempting to generate $\tlaplus$ proofs. The model appears to be drawing from its training on other formal systems, resulting in a hybrid syntax that combines elements from $\tlaplus$, Isabelle, and possibly other proof assistants. These syntax errors prevent the proof from being validated by $\TLAPM$, highlighting why syntactic correctness is a significant barrier to direct LLM-based proof generation for $\tlaplus$.

\section{Prompt Templates}
\label{appendix:prompt_templates}
\subsection{Prompt Template for Direct LLM Proof Generation}
\label{appendix:direct_single_pass_prompt}
\begin{lstlisting}[style=systemprompt]
You are an expert in TLA+ formal verification. Your task is to generate a complete, valid TLA+ proof for the given theorem.

Guidelines:
1. Generate a syntactically valid TLA+ proof using hierarchical proof structure with step numbers like <1>, <2>, etc.
2. Use proper TLA+ proof constructs: CASE, SUFFICES, TAKE, BY, etc.
3. Include necessary DEF references when using BY statements
4. Ensure all steps are properly justified
5. The proof should be complete and directly verifiable by TLAPM
6. DO NOT include any explanations or comments outside the TLA+ syntax
7. Return ONLY the complete TLA+ module with your proof integrated

Example of good proof structure:
```
THEOREM Example == \A x \in Nat: x + 0 = x
<1> SUFFICES ASSUME NEW x \in Nat
           PROVE  x + 0 = x
  OBVIOUS
<1>1 x + 0 = x BY SMT
<1> QED BY <1>1
```
\end{lstlisting}

\begin{lstlisting}[style=userprompt]
Here is a TLA+ module with a theorem that needs to be proved:

```
{tla_content}
```

Please generate a complete proof for the theorem '{theorem_name}' 

[IF_FAILED]((
Your previous proof attempt had the following issues when verified by TLAPM:

```
{feedback}
```

Please fix these issues and provide an improved proof that addresses these specific problems.
))

Return the entire TLA+ module with your proof integrated. The proof should be syntactically valid and verifiable by TLAPM.
\end{lstlisting}

\subsection{Prompt Template for LMGPA when Evaluating Math Benchmarks}
\label{appendix:template_prompt_for_decomposition}
There is no system prompt for LMGPA. The user prompt template is as follows:
\begin{lstlisting}[style=userprompt]
You are an expert specializing in decomposing complex TLA+
proof obligations into simpler sub-obligations. Your task is to
analyze this proof obligation and generate a logically sound
decomposition:

Format Instructions:
{format_instructions}

Context
{proof_context}

Goal:
{goal_obligation}

{{FEEDBACK_INFO}}

Follow these steps:
1. First, identify the key mathematical pattern or structure in this theorem
2. Express the transformation mathematically using TLA+ syntax, minimal drafts only
3. Break down the theorem into the simplest sub-claims that would establish the result
4. For each sub-claim, the final result should be ONLY in the following format:
   - A clear name
   - Necessary assumptions in TLA+ syntax
   - The precise hypothesis to prove
5. Provide an explanation of why the decomposition is valid, and why the new claims 
are easier to prove
6. Ensure your decomposition is sufficient to prove the original theorem,
and explain why.
7. For every proposed sub-claim, check if it is valid, and if not, provide an explanation of why it is not valid,
and how to fix it.
8. Try to fix the decomposition and subclaims until both the decomposition and subclaims are valid.
9. Write each of the simpler formula in a normalized form such that:
    - it has a name
    - it has a list of assumptions, where each assumption is either:
        - an expression, or
        - a definition identifier provided above
    - it has a hypothesis (goal) to prove, which is also a formula
    - PLEASE STRICTLY FOLLOW THE FORMAT INSTRUCTION
    - DON'T USE ENGLISH OR UNICODE. For logical symbols, use ASCII version, e.g. 
        - \A for \forall
        - \E for \exists
        - /\ for and
        - \/ for or
        - => for implication
        - <=> for iff
        - = for equality
        - /= for inequality
        - \in for set membership
        - Nat for natural number set
        - Int for integer set
        - Only +, -, *, % are allowed for arithmetic operations, division (/) and exponentiation (^) are not allowed
        - "NEW x \in Nat" or "NEW x \in Int" for adding new variables with domain, but this is only used in assumptions.
    - Every claim must be self-contained, that is, if there exists any free variables,
      then you need to add "NEW x \in Nat" or "NEW x \in Int" to the assumptions to specify the domain of that new variable
        - For example, if you generated a claim with 'Even(x)' as assumption,
          then you should add "NEW x \in Nat" to the assumptions to specify the domain of that new variable.
10. Double check the generated sub-claims, make sure there are no free variables left. Every variable used in assumptions and hypothesis should be defined as
   "NEW x \in Nat" in the assumptions.
11. Once done, conclude the sub-claims and return them in required format.

Guidelines:
- Use notation and syntax directly we mentioned above
- Limit explanations to 5-10 words per insight
- Focus on key mathematical properties and patterns (number theory properties, set relations, etc.)
- For each transformation, state the mathematical justification in <= 5 words
- Write each sub-claim in a normalized form with:
  - name='NAME' 
  - assumptions=['ASSUMPTION1', 'ASSUMPTION2', ...] 
  - hypothesis='GOAL'
- Use ASCII notation only for logical symbols (e.g., \A, \E, /\, \/, =>)
- Ensure all variables are properly quantified or declared
- Double-check for free variables

Here is a simple example of a normalized claim:
    name='L1' 
    assumptions=['NEW x \in Nat', 'NEW y \in Nat', '0 < y', 'y < x', 'x + y + (x * y) = 3'] 
    hypothesis='(x + 1) * (y + 1) = 4'
\end{lstlisting}

\section{Example Proofs Generated by Our System}
\label{appendix:example_proofs_found}

\begin{figure}[H]
    \centering
    \caption{Proofs generated by our LMGPA system from the mathematical benchmark suite.}
    \begin{minipage}[t]{0.46\textwidth}
        \centering
        \setlength{\abovecaptionskip}{0pt}
        \begin{lstlisting}[frame=single,style=tlaplusfancy]
---- MODULE mathd_numbertheory_234 ----
EXTENDS TLAPS, Integers

THEOREM Cube_Implies_N97_1 == 
\A N \in Nat : (N*N*N = 912673) => (N = 97)
OBVIOUS

THEOREM N97_Implies_Sum16_2 == 
ASSUME NEW a \in Nat, NEW b \in Nat, a >= 1, a <= 9, b <= 9, 10*a + b = 97
PROVE a + b = 16
OBVIOUS

THEOREM mathd_numbertheory_234 == 
\A a, b \in Nat : (a >= 1) /\ (a <= 9) /\ (b <= 9) /\ ((10 * a + b) * (10 * a + b) * (10 * a + b) = 912673) 
        => (a + b = 16)
BY Cube_Implies_N97_1, N97_Implies_Sum16_2
====\end{lstlisting}
    \end{minipage}
    \hfill
    \begin{minipage}[t]{0.46\textwidth}
        \centering
        \setlength{\abovecaptionskip}{0pt}
        \begin{lstlisting}[frame=single,style=tlaplusfancy]
---- MODULE amc12a_2002_p6 ----
EXTENDS TLAPS, Integers

THEOREM ExistenceOfM_1 == 
ASSUME NEW n \in Nat \ {0}
PROVE \E m \in Nat : m > n
OBVIOUS

THEOREM L1_2_1 == 
ASSUME NEW m \in Int
PROVE m * 1 <= m + 1
OBVIOUS

THEOREM ExistenceOfP_2 == 
ASSUME NEW m \in Nat
PROVE \E p \in Nat : m * p <= m + p
BY L1_2_1

THEOREM amc12a_2002_p6 == 
\A n \in Nat \ {0} : \E m \in Nat : 
    (m > n) /\ (\E p \in Nat : m * p <= m + p)
BY ExistenceOfM_1, ExistenceOfP_2
====\end{lstlisting}
    \end{minipage}

    \begin{minipage}[t]{0.46\textwidth}
        \centering
        \setlength{\abovecaptionskip}{0pt}
        \begin{lstlisting}[frame=single,style=tlaplusfancy]
---- MODULE exercise_1_1_4 ----
EXTENDS TLAPS, Integers

THEOREM DifferenceZero_1 == 
ASSUME NEW a \in Nat, NEW b \in Nat, NEW c \in Nat
PROVE (a*b)*c - a*(b*c) = 0
OBVIOUS

THEOREM ZeroInNat_2_1 == 0 \in Nat
OBVIOUS

THEOREM ZeroTimesAny_2_2 == 
ASSUME NEW n \in Int
PROVE 0 * n = 0
OBVIOUS

THEOREM ZeroMultiple_2 == 
ASSUME NEW n \in Nat
PROVE \E k \in Nat : 0 = k*n
BY ZeroInNat_2_1, ZeroTimesAny_2_2

THEOREM exercise_1_1_4 == 
\A a, b, c, n \in Nat : \E k \in Nat :
       (a * b) * c - a * (b * c) = k * n
BY DifferenceZero_1, ZeroMultiple_2 
====\end{lstlisting}
    \end{minipage}
    \hfill
    \begin{minipage}[t]{0.46\textwidth}
        \centering
        \setlength{\abovecaptionskip}{0pt}
        \begin{lstlisting}[frame=single,style=tlaplusfancy]
---- MODULE amc12_2000_p12 ----
EXTENDS Naturals, TLAPS

THEOREM Identity_1 == 
ASSUME NEW a \in Nat, NEW m \in Nat, NEW c \in Nat
PROVE (a+1)*(m+1)*(c+1) = a*m*c + a*m + a*c + m*c + a + m + c + 1
OBVIOUS

THEOREM MaxProduct_2 == 
ASSUME NEW a \in Nat, NEW m \in Nat, NEW c \in Nat, a + m + c = 12
PROVE (a+1)*(m+1)*(c+1) <= 125
OBVIOUS

THEOREM amc12_2000_p12 == 
\A a, m, c \in Nat : 
        (a + m + c = 12) => 
        (a * m * c + a * m + m * c + a * c <= 112)
BY Identity_1, MaxProduct_2
====\end{lstlisting}
    \end{minipage}
\end{figure}

\captionof{figure}{Proof generated by our LMGPA system from the distributed protocol benchmark suite.}
\begin{lstlisting}[frame=single,style=tlaplusfancy]
---- MODULE 9_sharded_kv ----
EXTENDS FiniteSetTheorems, TLAPS, TLC

CONSTANT Key
CONSTANT Value
CONSTANT Node

CONSTANT Nil

\* The key-value store state on each node.
VARIABLE table

\* The set of keys owned by each node.
VARIABLE owner

\* The set of active transfer messages.
VARIABLE transfer_msg

vars == <<table, owner, transfer_msg>>

Reshard(k,v,n_old,n_new) ==
    /\ table[n_old][k] = v
    /\ table' = [table EXCEPT ![n_old][k] = Nil]
    /\ owner' = [owner EXCEPT ![n_old] = owner[n_old] \ {k}]
    /\ transfer_msg' = transfer_msg \cup {<<n_new,k,v>>}

RecvTransferMsg(n, k, v) ==
    /\ <<n,k,v>> \in transfer_msg
    /\ transfer_msg' = transfer_msg \ {<<n,k,v>>}
    /\ table' = [table EXCEPT ![n][k] = v]
    \* Become the owner of this key.
    /\ owner' = [owner EXCEPT ![n] = owner[n] \cup {k}]

Put(n, k, v) == 
    /\ k \in owner[n]
    /\ table' = [table EXCEPT ![n][k] = v]
    /\ UNCHANGED <<owner, transfer_msg>>

Next == 
    \/ \E k \in Key, v \in Value, n_old,n_new \in Node : Reshard(k,v,n_old,n_new)
    \/ \E n \in Node, k \in Key, v \in Value : RecvTransferMsg(n,k,v)
    \/ \E n \in Node, k \in Key, v \in Value : Put(n,k,v)

Init == 
    /\ table = [n \in Node |-> [k \in Key |-> Nil]]
    \* Each node owns some subset of keys, and different nodes
    \* can't own the same key.
    /\ owner \in [Node -> SUBSET Key]
    /\ \A i,j \in Node : \A k \in Key : (k \in owner[i] /\ k \in owner[j]) => (i=j)
    /\ transfer_msg = {}

TypeOK ==
    /\ table \in [Node -> [Key -> Value \cup {Nil}]]
    /\ owner \in [Node -> SUBSET Key]
    /\ transfer_msg \in SUBSET (Node \times Key \times Value)

\* Keys unique.
Safety == 
    \A n1,n2 \in Node, k \in Key, v1,v2 \in Value : 
        (table[n1][k]=v1 /\ table[n2][k]=v2) => (n1=n2 /\ v1=v2)

Symmetry == Permutations(Key) \cup Permutations(Value) \cup Permutations(Node)

NextUnchanged == UNCHANGED vars


\* Inductive strengthening conjuncts
Inv238_1_0_def == \A NI \in Node : \A NJ \in Node : \A KI \in Key : \A VALI \in Value : ~(<<NI,KI,VALI>> \in transfer_msg) \/ (~(KI \in owner[NJ]))
Inv114_1_1_def == \A NJ \in Node : \A KI \in Key : (KI \in owner[NJ]) \/ ((table[NJ][KI] = Nil))
Inv1376_2_2_def == \A NI \in Node : \A NJ \in Node : \A KI \in Key : (NI = NJ /\ owner = owner) \/ (~(KI \in owner[NI])) \/ (~(KI \in owner[NJ]))
Inv1336_2_0_def == \A NI \in Node : \A NJ \in Node : \A KI \in Key : \A VALI \in Value : \A VALJ \in Value : (NI = NJ /\ owner = owner) \/ (~(<<NI,KI,VALJ>> \in transfer_msg) \/ (~(<<NJ,KI,VALI>> \in transfer_msg)))
Inv1476_2_1_def == \A NI \in Node : \A KI \in Key : \A VALI \in Value : \A VALJ \in Value : (VALI = VALJ /\ owner = owner) \/ (~(<<NI,KI,VALJ>> \in transfer_msg)) \/ (~(<<NI,KI,VALI>> \in transfer_msg))

\* The inductive invariant candidate.
IndAuto ==
  /\ TypeOK
  /\ Safety
  /\ Inv238_1_0_def
  /\ Inv114_1_1_def
  /\ Inv1376_2_2_def
  /\ Inv1336_2_0_def
  /\ Inv1476_2_1_def

ASSUME Fin == IsFiniteSet(Node) /\ IsFiniteSet(Key) /\ IsFiniteSet(Value)
ASSUME NilType == Nil \notin Node /\ Nil \notin Key /\ Nil \notin Value
ASSUME NodeNonEmpty == Node # {}


THEOREM Reshard_TypeOK_1_1_1 == 
ASSUME TypeOK, NEW k \in Key, NEW v \in Value, NEW n_old \in Node, NEW n_new \in Node, table[n_old][k] = v, table' = [table EXCEPT ![n_old][k] = Nil], owner' = [owner EXCEPT ![n_old] = owner[n_old] \ {k}], transfer_msg' = transfer_msg \cup {<<n_new,k,v>>}
PROVE TypeOK'
BY Fin, NodeNonEmpty, NilType DEF TypeOK

THEOREM RecvTransferMsg_TypeOK_1_1_2 == 
ASSUME TypeOK, NEW n \in Node, NEW k \in Key, NEW v \in Value, <<n,k,v>> \in transfer_msg, transfer_msg' = transfer_msg \ {<<n,k,v>>}, table' = [table EXCEPT ![n][k] = v], owner' = [owner EXCEPT ![n] = owner[n] \cup {k}]
PROVE TypeOK'
BY Fin, NodeNonEmpty, NilType DEF TypeOK

THEOREM Put_TypeOK_1_1_3 == 
ASSUME TypeOK, NEW n \in Node, NEW k \in Key, NEW v \in Value, k \in owner[n], table' = [table EXCEPT ![n][k] = v], UNCHANGED <<owner, transfer_msg>>
PROVE TypeOK'
BY Fin, NodeNonEmpty, NilType DEF TypeOK

THEOREM ReshardCase_1_1 == 
ASSUME IndAuto, NEW k \in Key, NEW v \in Value, NEW n_old \in Node, NEW n_new \in Node, Reshard(k,v,n_old,n_new)
PROVE TypeOK'
BY Reshard_TypeOK_1_1_1, RecvTransferMsg_TypeOK_1_1_2, Put_TypeOK_1_1_3 DEF Inv1376_2_2_def, Reshard, Inv1476_2_1_def, Inv1336_2_0_def, Inv114_1_1_def, Inv238_1_0_def, Safety, TypeOK, IndAuto

THEOREM RecvTransferMsgCase_1_2 == 
ASSUME IndAuto, NEW n \in Node, NEW k \in Key, NEW v \in Value, RecvTransferMsg(n,k,v)
PROVE TypeOK'
BY Fin, NodeNonEmpty, NilType DEF Inv1376_2_2_def, Inv1476_2_1_def, RecvTransferMsg, Inv1336_2_0_def, Inv114_1_1_def, Inv238_1_0_def, Safety, TypeOK, IndAuto

THEOREM PutCase_1_3 == 
ASSUME IndAuto, NEW n \in Node, NEW k \in Key, NEW v \in Value, Put(n,k,v)
PROVE TypeOK'
BY Fin, NodeNonEmpty, NilType DEF Inv1376_2_2_def, Inv1476_2_1_def, Inv1336_2_0_def, Inv114_1_1_def, Inv238_1_0_def, Safety, TypeOK, IndAuto, Put

THEOREM Reshard_Preserves_TypeOK_1 == 
ASSUME NEW k \in Key, NEW v \in Value, NEW n_old \in Node, NEW n_new \in Node, IndAuto, Reshard(k,v,n_old,n_new)
PROVE TypeOK'
BY ReshardCase_1_1, RecvTransferMsgCase_1_2, PutCase_1_3 DEF Inv1376_2_2_def, Reshard, Inv1476_2_1_def, Inv1336_2_0_def, Inv114_1_1_def, Inv238_1_0_def, Safety, TypeOK, IndAuto

THEOREM Reshard_Preserves_Safety_2 == 
ASSUME NEW k \in Key, NEW v \in Value, NEW n_old \in Node, NEW n_new \in Node, IndAuto, Reshard(k,v,n_old,n_new)
PROVE Safety'
BY Fin, NodeNonEmpty, NilType DEF Inv1376_2_2_def, Reshard, Inv1476_2_1_def, Inv1336_2_0_def, Inv114_1_1_def, Inv238_1_0_def, Safety, TypeOK, IndAuto

THEOREM Reshard_Preserves_Inv238_1_0_def_3 == 
ASSUME NEW k \in Key, NEW v \in Value, NEW n_old \in Node, NEW n_new \in Node, IndAuto, Reshard(k,v,n_old,n_new)
PROVE Inv238_1_0_def'
BY Fin, NodeNonEmpty, NilType DEF Reshard, Inv1376_2_2_def, Inv1476_2_1_def, Inv1336_2_0_def, Inv114_1_1_def, Inv238_1_0_def, Safety, TypeOK, IndAuto

THEOREM Reshard_Preserves_Inv114_1_1_def_4 == 
ASSUME NEW k \in Key, NEW v \in Value, NEW n_old \in Node, NEW n_new \in Node, IndAuto, Reshard(k,v,n_old,n_new)
PROVE Inv114_1_1_def'
BY Fin, NodeNonEmpty, NilType DEF Reshard, Inv1376_2_2_def, Inv1476_2_1_def, Inv1336_2_0_def, Inv114_1_1_def, Inv238_1_0_def, Safety, TypeOK, IndAuto

THEOREM Reshard_Preserves_Inv1376_2_2_def_5 == 
ASSUME NEW k \in Key, NEW v \in Value, NEW n_old \in Node, NEW n_new \in Node, IndAuto, Reshard(k,v,n_old,n_new)
PROVE Inv1376_2_2_def'
BY Fin, NodeNonEmpty, NilType DEF Inv1376_2_2_def, Reshard, Inv1476_2_1_def, Inv1336_2_0_def, Inv114_1_1_def, Inv238_1_0_def, Safety, TypeOK, IndAuto

THEOREM Reshard_Preserves_Inv1336_2_0_def_6 == 
ASSUME NEW k \in Key, NEW v \in Value, NEW n_old \in Node, NEW n_new \in Node, IndAuto, Reshard(k,v,n_old,n_new)
PROVE Inv1336_2_0_def'
BY Fin, NodeNonEmpty, NilType DEF Reshard, Inv1376_2_2_def, Inv1476_2_1_def, Inv1336_2_0_def, Inv114_1_1_def, Inv238_1_0_def, Safety, TypeOK, IndAuto

THEOREM Reshard_Preserves_Inv1476_2_1_def_7 == 
ASSUME NEW k \in Key, NEW v \in Value, NEW n_old \in Node, NEW n_new \in Node, IndAuto, Reshard(k,v,n_old,n_new)
PROVE Inv1476_2_1_def'
BY Fin, NodeNonEmpty, NilType DEF Reshard, Inv1376_2_2_def, Inv1476_2_1_def, Inv1336_2_0_def, Inv114_1_1_def, Inv238_1_0_def, Safety, TypeOK, IndAuto

THEOREM RecvTransferMsg_Preserves_TypeOK_8 == 
ASSUME NEW n \in Node, NEW k \in Key, NEW v \in Value, IndAuto, RecvTransferMsg(n,k,v)
PROVE TypeOK'
BY Fin, NodeNonEmpty, NilType DEF Inv1376_2_2_def, Inv1476_2_1_def, RecvTransferMsg, Inv1336_2_0_def, Inv114_1_1_def, Inv238_1_0_def, Safety, TypeOK, IndAuto

THEOREM RecvTransferMsg_Preserves_Safety_9 == 
ASSUME NEW n \in Node, NEW k \in Key, NEW v \in Value, IndAuto, RecvTransferMsg(n,k,v)
PROVE Safety'
BY Fin, NodeNonEmpty, NilType DEF Inv1376_2_2_def, Inv1476_2_1_def, RecvTransferMsg, Inv1336_2_0_def, Inv114_1_1_def, Inv238_1_0_def, Safety, TypeOK, IndAuto

THEOREM RecvTransferMsg_Preserves_Inv238_1_0_def_10 == 
ASSUME NEW n \in Node, NEW k \in Key, NEW v \in Value, IndAuto, RecvTransferMsg(n,k,v)
PROVE Inv238_1_0_def'
BY Fin, NodeNonEmpty, NilType DEF Inv1376_2_2_def, Inv1476_2_1_def, RecvTransferMsg, Inv1336_2_0_def, Inv114_1_1_def, Inv238_1_0_def, Safety, TypeOK, IndAuto

THEOREM RecvTransferMsg_Preserves_Inv114_1_1_def_11 == 
ASSUME NEW n \in Node, NEW k \in Key, NEW v \in Value, IndAuto, RecvTransferMsg(n,k,v)
PROVE Inv114_1_1_def'
BY Fin, NodeNonEmpty, NilType DEF Inv1376_2_2_def, Inv1476_2_1_def, RecvTransferMsg, Inv1336_2_0_def, Inv114_1_1_def, Inv238_1_0_def, Safety, TypeOK, IndAuto

THEOREM RecvTransferMsg_Preserves_Inv1376_2_2_def_12 == 
ASSUME NEW n \in Node, NEW k \in Key, NEW v \in Value, IndAuto, RecvTransferMsg(n,k,v)
PROVE Inv1376_2_2_def'
BY Fin, NodeNonEmpty, NilType DEF Inv1376_2_2_def, Inv1476_2_1_def, RecvTransferMsg, Inv1336_2_0_def, Inv114_1_1_def, Inv238_1_0_def, Safety, TypeOK, IndAuto

THEOREM RecvTransferMsg_Preserves_Inv1336_2_0_def_13 == 
ASSUME NEW n \in Node, NEW k \in Key, NEW v \in Value, IndAuto, RecvTransferMsg(n,k,v)
PROVE Inv1336_2_0_def'
BY Fin, NodeNonEmpty, NilType DEF Inv1376_2_2_def, Inv1476_2_1_def, RecvTransferMsg, Inv1336_2_0_def, Inv114_1_1_def, Inv238_1_0_def, Safety, TypeOK, IndAuto

THEOREM RecvTransferMsg_Preserves_Inv1476_2_1_def_14 == 
ASSUME NEW n \in Node, NEW k \in Key, NEW v \in Value, IndAuto, RecvTransferMsg(n,k,v)
PROVE Inv1476_2_1_def'
BY Fin, NodeNonEmpty, NilType DEF Inv1376_2_2_def, Inv1476_2_1_def, RecvTransferMsg, Inv1336_2_0_def, Inv114_1_1_def, Inv238_1_0_def, Safety, TypeOK, IndAuto

THEOREM Put_Preserves_TypeOK_15 == 
ASSUME NEW n \in Node, NEW k \in Key, NEW v \in Value, IndAuto, Put(n,k,v)
PROVE TypeOK'
BY Fin, NodeNonEmpty, NilType DEF Inv1376_2_2_def, Inv1476_2_1_def, Inv1336_2_0_def, Inv114_1_1_def, Inv238_1_0_def, Safety, TypeOK, IndAuto, Put

THEOREM Put_Preserves_Safety_16 == 
ASSUME NEW n \in Node, NEW k \in Key, NEW v \in Value, IndAuto, Put(n,k,v)
PROVE Safety'
BY Fin, NodeNonEmpty, NilType DEF Inv1376_2_2_def, Inv1476_2_1_def, Inv1336_2_0_def, Inv114_1_1_def, Inv238_1_0_def, Safety, TypeOK, IndAuto, Put

THEOREM Put_Preserves_Inv238_1_0_def_17 == 
ASSUME NEW n \in Node, NEW k \in Key, NEW v \in Value, IndAuto, Put(n,k,v)
PROVE Inv238_1_0_def'
BY Fin, NodeNonEmpty, NilType DEF Inv1376_2_2_def, Inv1476_2_1_def, Inv1336_2_0_def, Inv114_1_1_def, Inv238_1_0_def, Safety, TypeOK, IndAuto, Put

THEOREM Put_Preserves_Inv114_1_1_def_18 == 
ASSUME NEW n \in Node, NEW k \in Key, NEW v \in Value, IndAuto, Put(n,k,v)
PROVE Inv114_1_1_def'
BY Fin, NodeNonEmpty, NilType DEF Inv1376_2_2_def, Inv1476_2_1_def, Inv1336_2_0_def, Inv114_1_1_def, Inv238_1_0_def, Safety, TypeOK, IndAuto, Put

THEOREM Put_Preserves_Inv1376_2_2_def_19 == 
ASSUME NEW n \in Node, NEW k \in Key, NEW v \in Value, IndAuto, Put(n,k,v)
PROVE Inv1376_2_2_def'
BY Fin, NodeNonEmpty, NilType DEF Inv1376_2_2_def, Inv1476_2_1_def, Inv1336_2_0_def, Inv114_1_1_def, Inv238_1_0_def, Safety, TypeOK, IndAuto, Put

THEOREM Put_Preserves_Inv1336_2_0_def_20 == 
ASSUME NEW n \in Node, NEW k \in Key, NEW v \in Value, IndAuto, Put(n,k,v)
PROVE Inv1336_2_0_def'
BY Fin, NodeNonEmpty, NilType DEF Inv1376_2_2_def, Inv1476_2_1_def, Inv1336_2_0_def, Inv114_1_1_def, Inv238_1_0_def, Safety, TypeOK, IndAuto, Put

THEOREM Put_Preserves_Inv1476_2_1_def_21 == 
ASSUME NEW n \in Node, NEW k \in Key, NEW v \in Value, IndAuto, Put(n,k,v)
PROVE Inv1476_2_1_def'
BY Fin, NodeNonEmpty, NilType DEF Inv1376_2_2_def, Inv1476_2_1_def, Inv1336_2_0_def, Inv114_1_1_def, Inv238_1_0_def, Safety, TypeOK, IndAuto, Put

THEOREM Inductiveness == 
IndAuto /\ Next => IndAuto'
BY Reshard_Preserves_TypeOK_1, Reshard_Preserves_Safety_2, Reshard_Preserves_Inv238_1_0_def_3, Reshard_Preserves_Inv114_1_1_def_4, Reshard_Preserves_Inv1376_2_2_def_5, Reshard_Preserves_Inv1336_2_0_def_6, Reshard_Preserves_Inv1476_2_1_def_7, RecvTransferMsg_Preserves_TypeOK_8, RecvTransferMsg_Preserves_Safety_9, RecvTransferMsg_Preserves_Inv238_1_0_def_10, RecvTransferMsg_Preserves_Inv114_1_1_def_11, RecvTransferMsg_Preserves_Inv1376_2_2_def_12, RecvTransferMsg_Preserves_Inv1336_2_0_def_13, RecvTransferMsg_Preserves_Inv1476_2_1_def_14, Put_Preserves_TypeOK_15, Put_Preserves_Safety_16, Put_Preserves_Inv238_1_0_def_17, Put_Preserves_Inv114_1_1_def_18, Put_Preserves_Inv1376_2_2_def_19, Put_Preserves_Inv1336_2_0_def_20, Put_Preserves_Inv1476_2_1_def_21 DEF Reshard, Inv1376_2_2_def, Inv1476_2_1_def, RecvTransferMsg, Inv1336_2_0_def, Inv114_1_1_def, Inv238_1_0_def, Safety, TypeOK, IndAuto, Next, Put
====\end{lstlisting}

    \fi
\end{document}